\documentclass[useAMS,usegraphicx, usenatbib]{mn2e}
\usepackage{aas_macros}
\usepackage{times}
\usepackage{aas_macros}
\usepackage{footnote} 
\usepackage{verbatim, amsmath, amsfonts, amssymb,amssymb}
\usepackage[utf8x]{inputenc}
\usepackage{setspace}
\usepackage{psfrag}
\usepackage{graphicx,subfigure}
\usepackage{multirow}
\makesavenoteenv{tabular}
\usepackage{rotating}
\usepackage[usenames,dvipsnames,table,svgnames,rgb,table]{xcolor}
\usepackage{booktabs, longtable}
\usepackage{tikz}
\usetikzlibrary{matrix}
\usepackage{booktabs}
\usepackage{pgfplots}
\usepackage{pgfplotstable}
\usepackage{multicol}
\usepackage{listings}
\usepackage{mathrsfs}
\usepackage{enumitem}
\usepackage{enumerate}
\usepackage{float}

\begin{document}
\pagestyle{myheadings}
\title{Probing the stellar   initial mass function with high-\emph{z} supernovae}
\markboth{High-\emph{z} initial mass function}{}

\author[R. S. de Souza, E. E. O. Ishida, D. J., Whalen, J. Johnson,  A. Ferrara]
{R. S. de Souza $^{1,2}\thanks{e-mail: rafael.2706@gmail.com (RSS)}$;
E. E. O. Ishida  $^{3,4}$;
D. J. Whalen $^{5,6}$;
J.  L. Johnson $^{5}$;
A. Ferrara$^{7}$\\
\\
$^{1}$Korea Astronomy \& Space Science Institute, Daejeon 305-348, Korea\\
$^{2}$MTA E\"otv\"os University,  EIRSA "Lendulet" Astrophysics Research Group, Budapest 1117, Hungary\\
$^{3}$Max-Planck-Institut f\"ur Astrophysik, Karl-Schwarzschild-Str. 1, D-85748 Garching, Germany\\
$^{4}$IAG, Universidade de S\~ao Paulo, Rua do Mat\~ao 1226, 05508-900, S\~ao Paulo,  Brazil\\
$^{5}$Los Alamos National Laboratory, Los Alamos, NM 87545, USA\\
$^{6}$Universit\"at Heidelberg, Zentrum f\"ur Astronomie, Institut f\"ur Theoretische Astrophysik, 
Albert-Ueberle-Str. 2, 69120 Heidelberg, Germany\\
$^{7}$Scuola Normale Superiore, Piazza dei Cavalieri 7, 56126 Pisa, Italy
}

 \date{Accepted -- Received  --} 

\pagerange{\pageref{firstpage}--\pageref{lastpage}} \pubyear{2010}

\maketitle
\label{firstpage}

%------------------------------------------------------------------------------------------------------------------------------------------------------------------------------------------------------------------------%

\begin{abstract}
 
The first supernovae will soon be visible at the edge of the observable universe, revealing the birthplaces of  Population III stars.  With upcoming near-infrared missions, 
a broad analysis of the detectability of high-\emph{z} supernovae is paramount.  We combine cosmological and radiation transport simulations, instrument specifications,
and survey strategies to create synthetic observations of primeval core-collapse, Type IIn and pair-instability supernovae with the {\it James Webb Space Telescope} (\textit{JWST}).  We 
show that a dedicated observational campaign with the 
\textit{JWST} can detect up to $\sim 15$ pair-instability explosions, $\sim 300$ core-collapse 
supernovae, but less than one Type IIn explosion per year, depending on the Population III star formation history.  Our synthetic survey also shows that $\approx 1-2 \times10^2$ 
supernovae detections, depending on the accuracy of  the classification,  are sufficient to discriminate between a Salpeter and flat mass distribution for high redshift  stars  with a confidence level greater 
than 99.5 per cent.   We discuss how the purity of the sample affects our results and how supervised learning methods may help  to  discriminate   between CC and PI SNe. 

\end{abstract}

%------------------------------------------------------------------------------------------------------------------------------------------------------------------------------------------------------------------------%

\section{Introduction}

The emergence of Population III (Pop III) stars ended the cosmic dark ages and began the production of elements heavier than lithium \citep{Karlsson2013}. But because 
Pop III stars have never been observed, their properties remain uncertain.  Even detections of Pop III stars through strong gravitational lensing are very unlikely \citep{
Rydberg2013}, so it may be decades before they are directly observed.  The most viable strategy for probing the properties of the first stars is direct detection of their 
gamma-ray bursts \citep[GRBs;][]{Bromm2006,desouza2011a,desouza2011b,Campisi2011,Nagakura2012,Mesler2013a} and supernovae \citep[SNe;][]{kasen2011,
desouza2013b,Whalen2013a}. Pop III SNe may soon be found in deep-field surveys by the {\it James Webb Space Telescope} ({\it JWST})\footnote{http://www.jwst.nasa.gov} 
and all-sky surveys by \textit{Euclid}\footnote{http://www.euclid-ec.org/} \citep[e.g.,][]{Whalen2012b,Johnson2013b}, providing unprecedented insights into cosmic evolution 
\citep{Bromm2013} and the properties of primordial mini-haloes \citep[e.g.,][]{whalen2008,Biffi2013,desouza2013a,desouza2013c}.  

The \textit{JWST} will be an infrared-optimized space telescope with four instruments, the Near-Infrared Camera (NIRCam), the Near-Infrared Spectrograph,  
the near-infrared Tunable Filter Imager  and the Mid Infrared Instrument, with enough fuel for a 10-yr mission \citep{Gardner2006}.  Our analysis is based on 
NIRCam photometry in the 0.6-5$\mu$m bands, one of whose science goals is the exploration of the end of the dark ages.  Complementary all-sky searches will also be 
possible with other NIR missions such as the {\it Wide-Field Infrared Survey Telescope (WFIRST)},  the {\it Wide-field Imaging Surveyor for High-redshift (WISH)} and \textit{Euclid}.  
Pop III SNe will also be visible in the radio at 21 cm to the \textit{Expanded Very Large Array} (eVLA), \textit{eMERLIN} and future radio facilities like the \textit{Square 
Kilometre Array} \citep{Meiksin2013}.

The quest for the first stars has been further motivated by the discovery of pair-instability (PI) SN candidates SN 2007bi at $z =$ 0.127 \citep{GalYam2009,Young2010} and
SN 2213-1745 at $z =$ 2.06 \citep{Cooke12} and the pulsational pair-instability (PPI) SN candidate SN 1000$+$0216 at $z =$ 3.90 \citep{Cooke12}.  These lower redshifts 
are less favorable to the formation of massive progenitors than in the early universe.  There is also evidence that some Pop III stars may be less massive than suggested by 
previous studies \citep[][]{abel2002,omukai2003,yoshida2006}.  Observationally, there is evidence of 15 - 50 $M_{\bigodot}$ Pop III stars in the fossil abundance record, the 
ashes of early SNe thought to be imprinted on ancient metal-poor stars \citep{Beers2005,Frebel2005,Cayrel2004,Iwamoto2005,Lai2008,Joggerst2010,Caffau2012}. Recent 
numerical simulations have produced Pop III stars with masses below 50 $M_{\bigodot}$ \citep{greif2011,clark2011,hosokawa2011,Stacy2012a}, and in some cases as low 
as $\sim 1 -5 M_{\bigodot}$ \citep{Stacy2013}. A number  of such stars could still be living today.  

Pop III stars from $15-40 M_{\bigodot}$ die as core-collapse (CC) SNe and 85-260 $M_{\bigodot}$ stars explode as far more energetic PI SNe \citep{heger2002,
chatzopoulos2012}.   If a CC SN shock wave collides with a dense circumstellar shell ejected in an outburst a few years prior to the death of the star, 
a so-called Type IIn SN \citep{Whalen2013d}.  Detection limits in redshift for PI SNe \citep{Weinmann2005,wise2005,pan2012,Hummel2012,Whalen2013a,Whalen2013b,desouza2013b}, CC SNe \citep{Mesinger2006,Whalen2013c} and Type IIn \citep{Tanaka2012,Tanaka2013,
Whalen2013d} have shown that these events will be visible in the NIR at $z \sim$ 10 - 30.  There is a great expectation that the hunt for the first stars will soon be prosperous, 
making our study timely and in synergy with the established literature.

The observable properties of a galaxy, such as color and magnitude, are mainly determined by the initial mass function (IMF) and formation history of its stars. These serve 
as essential inputs, from which semi-analytical models of galaxy evolution can predict colors and luminosities \citep[e.g., ][]{Bastian2010}.  Probing the primordial IMF will not
only improve our understanding of the early stages of star formation, but also help to uncloak a long lasting puzzle:  is the IMF universal or does it evolve over cosmic time?  
We show that the number of primordial SN detections required to reveal the Pop III IMF is feasible.  Such an achievement would be one of the main, ground breaking 
accomplishments of extragalactic astrophysics in the next decade. 
 
Here, we extend the work of \citet[][hereafter DS13]{desouza2013b} in several ways: (i) we include spectral energy distributions (SEDs) for CC and Type IIn SNe, and new
PI SN SEDs calculated with improved physics, in particular a better implementation of self-gravity into {\sc rage} code,   with a reran  of the the red  and  blue supergiant models \citep{Whalen2013a,Whalen2013b};\footnote{The original models were brighter because self-gravity was not 
properly implemented in {\sc rage}, and the shocks broke out of the star
with too much kinetic energy.  We have confirmed that this did not affect any of 
the conclusions from DS13 on the 
detectability of these events because the new explosions are still 
quite visible at z $\sim$ 20.  The peak luminosities in our newer light 
curves are in good agreement with those of \citet{kasen2011,Dessart2013}.
} (ii) we analyze how observational strategies affect the detection rate of each SN type; (iii) we construct a statistical study 
of the feasibility of constraining the Pop III IMF, and (iv) we present the first application of photometric classification techniques to study how well can we discriminate between different Pop III SNe types.   The outline of the rest of this paper is as follows.  In Section \ref{sec:cosmo_sim}, we discuss our cosmological simulations, 
and in Section \ref{sec:SED} we review our SN light curve (LC) calculations.  Survey strategies and our synthetic observations are described in Section \ref{sec:survey}. In 
Section \ref{sec:IMF} we discuss future constraints on the Pop III IMF and our adopted statistics. In Section \ref{sec:class} we discuss how to identify each SN type and how this affects our results, and in Section \ref{sec:conclusions} we present our conclusions.

%------------------------------------------------------------------------------------------------------------------------------------------------------------------------------------------------------------------------%

\section{Cosmological simulations}
\label{sec:cosmo_sim}

The SN rate is directly related to the cosmic star formation rate (SFR), which largely depends on the ability of primordial gas to cool and condense.  Hydrogen molecules 
($H_2$) are the primary coolant in primordial gas clouds, and they are sensitive to the soft ultraviolet background (UVB).  The UVB in the $H_2$-disassociating Lyman-Werner 
(LW) bands delays star formation inside mini-haloes, and self-consistent cosmological simulations are required to properly account for this effect.  We derive cosmic 
SFRs from a cosmological N-body, hydro and chemistry simulation \citep{Johnson2013a} done with a modified version \citep{Schaye2010} of the smoothed-particle 
hydrodynamics (SPH) code \textsc{gadget} \citep{Springel2001,Springel2005}.  The modifications include line cooling in photoionization equilibrium for 11 elements 
\citep[H, He, C, N, O, Ne, Mg, Si, S, Ca, Fe; ][]{Wiersma2009}, prescriptions for SN mechanical feedback and metal enrichment, a non-equilibrium primordial chemistry 
network, and molecular cooling by $H_2$ and HD \citep{Abel1997,Galli1998,yoshida2006,Maio2007}. The transition from Pop III to Pop II/I SF occurs when the metallicity
of the gas in the simulation, $Z$, exceeds the critical value $Z_{crit}= 10^{-4} Z_{\bigodot}$ \citep{omukai2001,Bromm2001,Schneider2002,Mackey2003,Tornatore2007,
maio2010}.  The simulation starts from cosmological initial conditions and has a cubic volume 4 Mpc (comoving) on a side with periodic boundary conditions.  It includes 
both dark matter (DM) and gas, with SPH gas and DM particle masses of $1.25 \times 10^{3} M_{\bigodot}$ and $6.16 \times 10^{3} M_{\bigodot}$, respectively.  The 
simulation is launched with an equal number of SPH and DM particles, $684^3$.  We adopt the WMAP7 cosmological parameters \citep{Komatsu2009}:  $\Omega_{\rm 
m} = 0.265$, $\Omega_{\rm b} = 0.0448$, $\Omega_{\rm \Lambda} = 0.735$, $H_0 = 71 \rm km~s^{-1} \rm Mpc^{-1}$, and $\sigma_8 = 0.81$.

We assume a Salpeter IMF for Pop III stars \citep{Salpeter1955}, with upper and lower limits $M_{\rm upper} = 500 M_{\bigodot}$ and $M_{\rm lower} = 21 M_{\bigodot}
$, respectively \citep{Bromm2004,Karlsson2008}.  The prescriptions for Pop III stellar evolution and chemical feedback track the enrichment of gas by the 11 elements 
listed above individually \citep{Maio2007}, using nucleosynthetic yields from the explosions of metal-free stars \citep{heger2002,Heger2010}. We include a LW background 
from proximate sources and sources outside the simulation volume.  The photodissociation rate due to this background is included in the reaction network throughout the 
run.  In addition to the LW background, strong spatial and temporal variations in the LW flux can be produced locally by individual stellar sources \citep{Dijkstra2008,
Ahn2009}, and this effect is included by summing the local LW flux contributions from all star particles.  Self-shielding of $H_2$ to the LW background was implemented 
with shielding functions from \citet{Wolcott2011}.

\subsection{SN Rate}
\label{sec:SFH}

Since massive stars have cosmologically short lifetimes, $\approx 30(M/8M_{\bigodot})^{-2.5}$ Myr, it is usually assumed that the SN rate, $\dot{n}_{\rm SN}$, traces the 
cosmic SFR with redshift.  This rate can be derived for a particular mass range (and hence explosion type) from the IMF by 
\begin{equation}
\dot{n}_{\rm SN}(z) = SFR(z)\frac{\int^{M_{\rm max}}_{M_{\rm min}}\psi(M)dM}{\int^{500 M_{\bigodot}}_{15 M_{\bigodot}}M\psi(M)dM}\quad\rm yr^{-1}Mpc^{-3},
\label{eq:snrate}
\end{equation}
where $\psi(M)$ is the Pop III IMF, which we assume to have an either Salpeter ($dN/dLog M \propto M^{-1.35}$) or flat ($dN/dLog M \propto M$) slope.  In reality, any 
IMF other than Salpeter is not entirely self-consistent, since mechanical feedback by SNe, metal enrichment, and the LW background can be dependent on the chosen 
IMF, and these in turn feed back into the SFR.  But since Pop II SFRs are much higher than Pop III rates at z $<$ 15, the contribution to feedback from Pop III stars is 
minor at later epochs.  The overall SFR is therefore not very sensitive to the Pop III IMF, except perhaps at z $>$ 15. Nevertheless, since most of the SNe occur at lower 
redshifts, this error is smaller than other astrophysical and numerical uncertainties and is thus subsumed in the wide range of parameters explored here.  

The lower and upper limits for CC and PI SNe are $M_{min}=$ 15 and 140 $M_{\bigodot}$ and $M_{\rm max} =$ 40 and 260 $M_{\bigodot}$, respectively.  New results 
from \cite{chatzopoulos2012} show that Pop III stars down to $\sim 85 M_{\bigodot}$ may produce PI SNe if they are rapidly rotating, which could increase the PI SN rate 
by a factor of $\sim 4$.  Our flat IMF is motivated by recent cosmological simulations of primordial star formation \citep{greif2011,clark2011,hirano13}. The fraction of CC 
SNe that are Type IIn is not well determined but is assumed to be $10^{-3}$ of all CC SNe.  This fraction is in agreement with \citet{Tanaka2012} and current constraints 
from the \textit{Robotic Optical Transient Search Experiment-IIIb} \citep{Quimby2013}.  

\begin{figure}
\includegraphics[width=1.025\columnwidth]{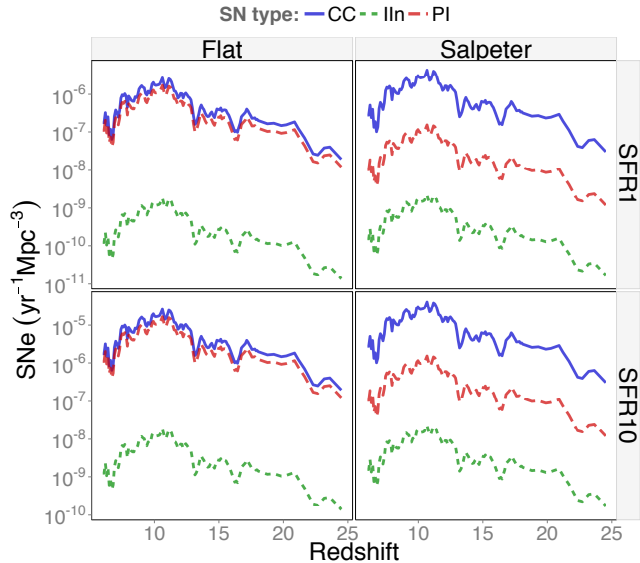}
\caption{$\dot{n}_{\rm SN}(z)$ for CC, IIn and PI SNe for the two IMFs (columns) and SFRs (rows) as a function of redshift.}
\label{fig:SNe_rate}
\end{figure}

\begin{figure}
\includegraphics[width=1\columnwidth]{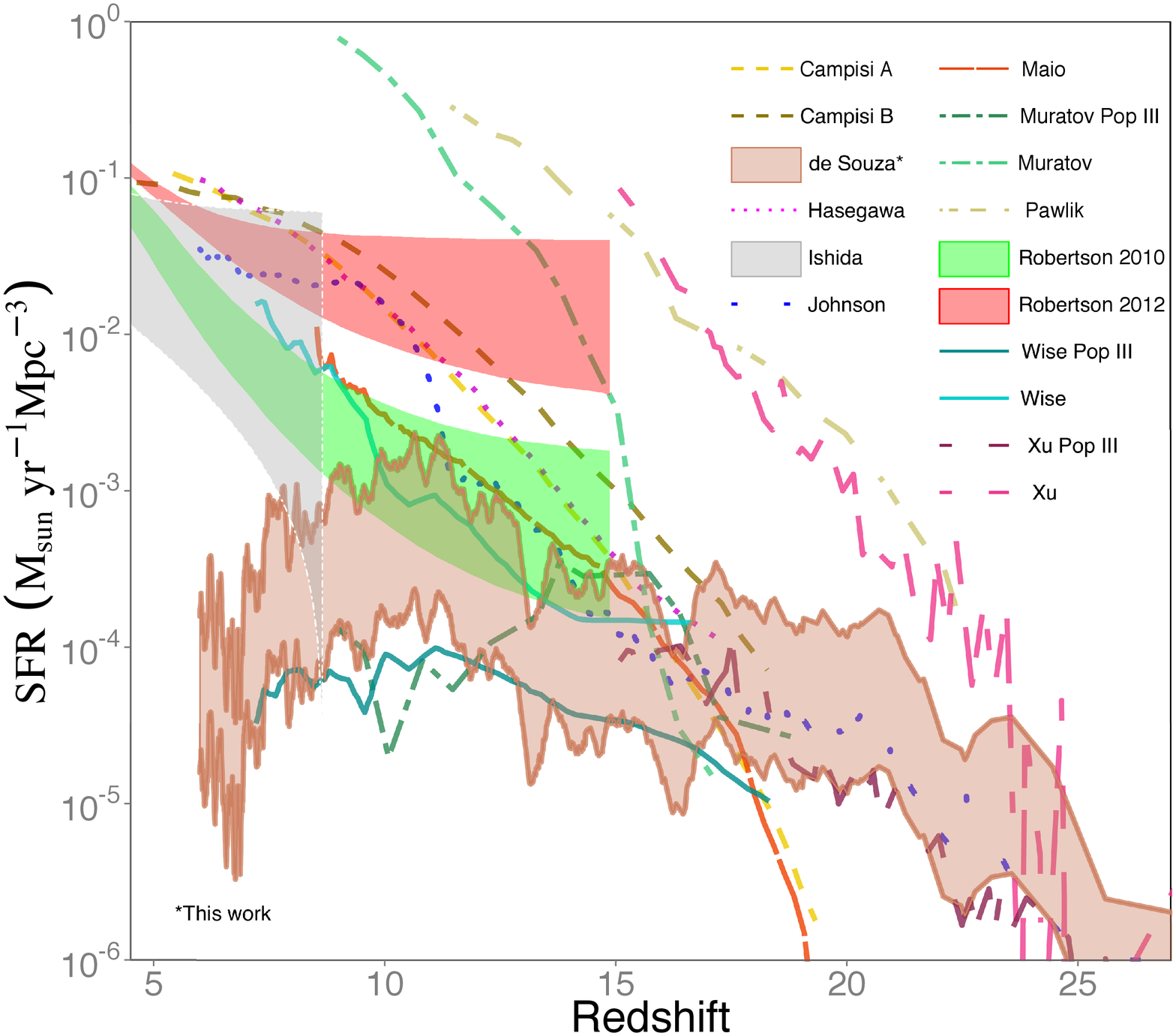}
\caption{Cosmic SFRs as a function of redshift.  The grey and red   bands are rates inferred from GRBs \citep{Ishida2011,Robertson2012} and the green band 
are rates inferred from UV-selected galaxies  \citep{Robertson2010}.  The SFRs compiled from simulations are from \citet{Campisi2011,maio2011,Wise2012,Hasegawa2013,Johnson2013a,Muratov2013,Pawlik2013,Xu2013} \citep[see ][for other SFRs that fall within those shown here]{Tornatore2007,
Trenti2009}.  The two lower and upper limits from the salmon band represent  the Pop IIII SFR models SFR1 and SFR10 from this  work \citep[see also]{desouza2013b}.   Boundaries from de Souza, Muratov Pop III, Wise Pop III and Xu Pop III SFR are for Pop III stars only, the others are Pop II $+$ I SFRs.}
\label{fig:SFRIII}
\end{figure}

Fig.  \ref{fig:SNe_rate} shows $\dot{n}_{\rm SN}$ from CC, Type IIn and PI SNe for different IMFs and SFRs separately.  Model SFR1 refers to the standard Pop III SFR, adopted in our previous work \citep[DS13,][]{
Johnson2013a}, while model SFR10 is a more optimistic case with an Pop III SFR that is ten times higher, to account for numerical and astrophysical uncertainties.  Our wide 
range of parameters aims to bracket the SFRs in the literature (Fig. \ref{fig:SFRIII}, see also Fig. 3 from \citealt{Whalen2013e}).  These rates come from numerical 
simulations \citep[][see \citealt{Tornatore2007,Trenti2009} for other SFRs that fall within the range shown here]{Campisi2011,maio2011,Wise2012,Hasegawa2013,Johnson2013a,Muratov2013,Pawlik2013,Xu2013}, long-duration GRB rates \citep{Ishida2011,Robertson2012} and  UV-selected galaxies  \citep{Robertson2010}.  The 
rates predicted by both observations and simulations vary by about two orders of magnitude from $z\sim 10-25$.  

The Pop III SFRs in most of the models are taken from those of a few protogalaxies in their simulation volumes and are therefore subject to small box statistics. At high redshift, 
galactic halos are rare and correspond to large peaks in the Gaussian probability distribution of initial fluctuations \citep{Barkana2004}.  In simulations, periodic boundary 
conditions are usually assumed, thereby forcing the mean density of the box to be the cosmic mean density. This excludes density modes with wavelengths longer than 
the box size (4 Mpc in our case) from the simulations and results in an underestimate of the mean number of rare, biased halos.  Accounting for these missing modes 
\citep{Barkana2004}, we estimate that the true SFR at $z = $ 7 (20) could be larger by a factor of $\sim$ 1.3 (7) if star formation is dominated by atomically cooled haloes, 
or a factor of $\sim$ 1.1 (2) if star-formation is dominated by haloes cooled by H$_2$.  The simulations also include a variety of feedback processes that lead to a wide 
range of evolution in stellar populations over redshift.   

%------------------------------------------------------------------------------------------------------------------------------------------------------------------------------------------------------------------------%

%------------------------------------------------------------------------------------------------------------------------------------------------------------------------------------------------------------------------%

%------------------------------------------------------------------------------------------------------------------------------------------------------------------------------------------------------------------------%

\section{SN LC models }
\label{sec:SED}

The SN spectra in our study are calculated in three steps.  First, the Pop III star is evolved from the zero-age main sequence to central collapse and then explosion in the 
one-dimensional (1D) {\sc kepler} stellar evolution code \citep{Weaver1978,Woosley2002}.  Nuclear burning is calculated with a 19-isotope network until oxygen depletion 
in the core, and with a 128-isotope quasi-equilibrium network thereafter.  The {\sc kepler} simulation is halted after the end of explosive burning, when the shock is still deep
inside the star.  The blast profile, surrounding star, and stellar envelope (a low-density $r^{-2}$ wind profile) are then mapped onto a 1D spherical grid in the Los Alamos
{\sc rage} code \citep{Gittings2008} and evolved through shock breakout from the star and subsequent expansion into the intergalactic medium (IGM).  Finally, {\sc rage} profiles 
are post processed with the Los Alamos {\sc spectrum} code \citep{Frey2013} and OPLIB opacities\footnote{http://rdc.llnl.gov}\citep{oplib} 
to obtain spectra at 14899 wavelengths.  We summarize the properties of our SNe in Table 1.

\paragraph*{PI SNe} 

We consider 150, 175, 200, 225, and 250 $M_{\bigodot}$ PI SN progenitors at zero-metallicity (z-series) and $Z =$ $10^{-4} Z_{\bigodot}$ (u-series) \citep{Joggerst2011,
Whalen2013b}.  The u-series stars die as red supergiants and the z-series stars die as compact blue giants.  It is generally thought that most stars in this mass range die 
as red supergiants due to convective mixing over their lives, even at zero metallicity. The onset of the PI and explosion is an emergent feature of the stellar evolution 
model and does not have to be artificially triggered.  PI SNe exhibit little internal mixing that would break the spherical symmetry of the star during the explosion and are 
generally well-described by 1D models.  They make up to 40 $M_{\bigodot}$ of $\ensuremath{^{56}\mathrm{Ni}}$ and release up to 10$^{53}$ erg, which powers the LC 
at intermediate to late times.  We evolve these explosions out to 3 yrs because they can be bright out to these times, in part because of the longer radiation diffusion 
timescales in their massive ejecta. The red supergiants and blue giants in our grid of models bracket the range of structures expected for these stars.  Their actual 
structures, and hence explosion LCs, may be intermediate to those here.
 
\paragraph*{CC SNe} 

The CC SNe in our study are $15-40 M_{\bigodot}$, with explosion energies, $E$, of 0.6, 1.2, and 2.4 $\times 10^{51}$ erg (the B, D, and G models, respectively).  They
have the same metallicities as our PI SN models and also die as red supergiants (z-series) and compact blue giants (u-series), a total of 18 models.  Unlike the PI SNe
described above, these explosions must be artificially triggered with linear momentum pistons in {\sc kepler}.  Before mapping them to {\sc rage}, these SNe are evolved from 
the end of explosive burning until just before shock breakout in two dimensions (2D) in the {\sc castro} adaptive mesh refinement (AMR) code \citep{Almgren2010}.  We do 
this to capture the violent mixing between elemental shells in the star before the SN shock ruptures its surface, as described in \citet{Joggerst2010}. These 2D profiles are 
then averaged in angle, mapped into {\sc rage}, and evolved through shock breakout out to 4 months.  Our procedure approximates the order in which absorption and 
emission lines due to various elements might appear in the 1D spectra over time because of mixing.  The structures of these stars again span those expected for Pop III 
stars in this mass range.

%------------------------------------------------------------------------------------------------------------------------------------------------------------------------------------------------------------------------%

\begin{figure}
\centering
\includegraphics[scale=0.35]{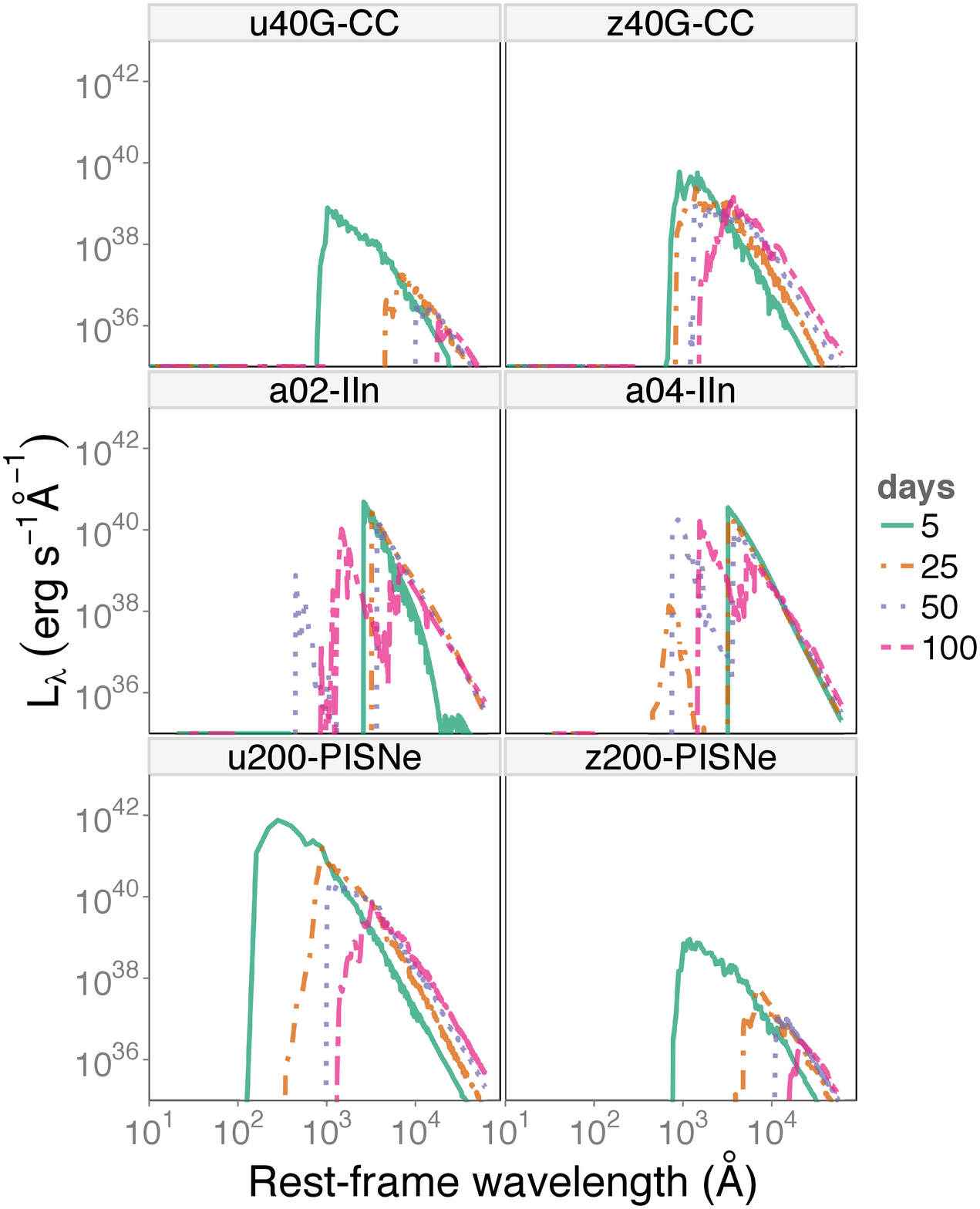}
\caption{Rest-frame spectral evolution of the u40G and z40G CC SNe, the a02 and a04 Type IIne, and the u200 and z200 PI SNe.  Fireball spectra are at $\approx$ 5 days 
(green solid line), 25 days (orange dot long-dashed line), 50 days (violet dotted line), and 100 days (pink dot-dashed line). }
\label{fig:SNezoo}
\end{figure}

%------------------------------------------------------------------------------------------------------------------------------------------------------------------------------------------------------------------------%

\paragraph*{IIn} 

As noted in the Introduction, Type IIn SNe occur when ejecta from a CC explosion crash into a dense shell ejected by the star prior to death.  The collision lights up the 
shell in the UV in the rest frame, whose spectral peak is then redshifted in the NIR at high-$z$.  Such events can be super-luminous because of the large radius of the 
shell on impact, which can be $\sim$ 10$^{16}$ cm.  The star can range in mass from 20 - 40 $M_{\bigodot}$, but we adopt the z40G model as our fiducial CC SN and 
then consider collisions with five shells whose masses are 0.1 - 20 $M_{\bigodot}$.  These explosions were evolved out to 500 days, past breakout of the shock through 
the outer surface of the shell.  

\paragraph*{}

\begin{figure*}
\centering
\includegraphics[scale=0.5]{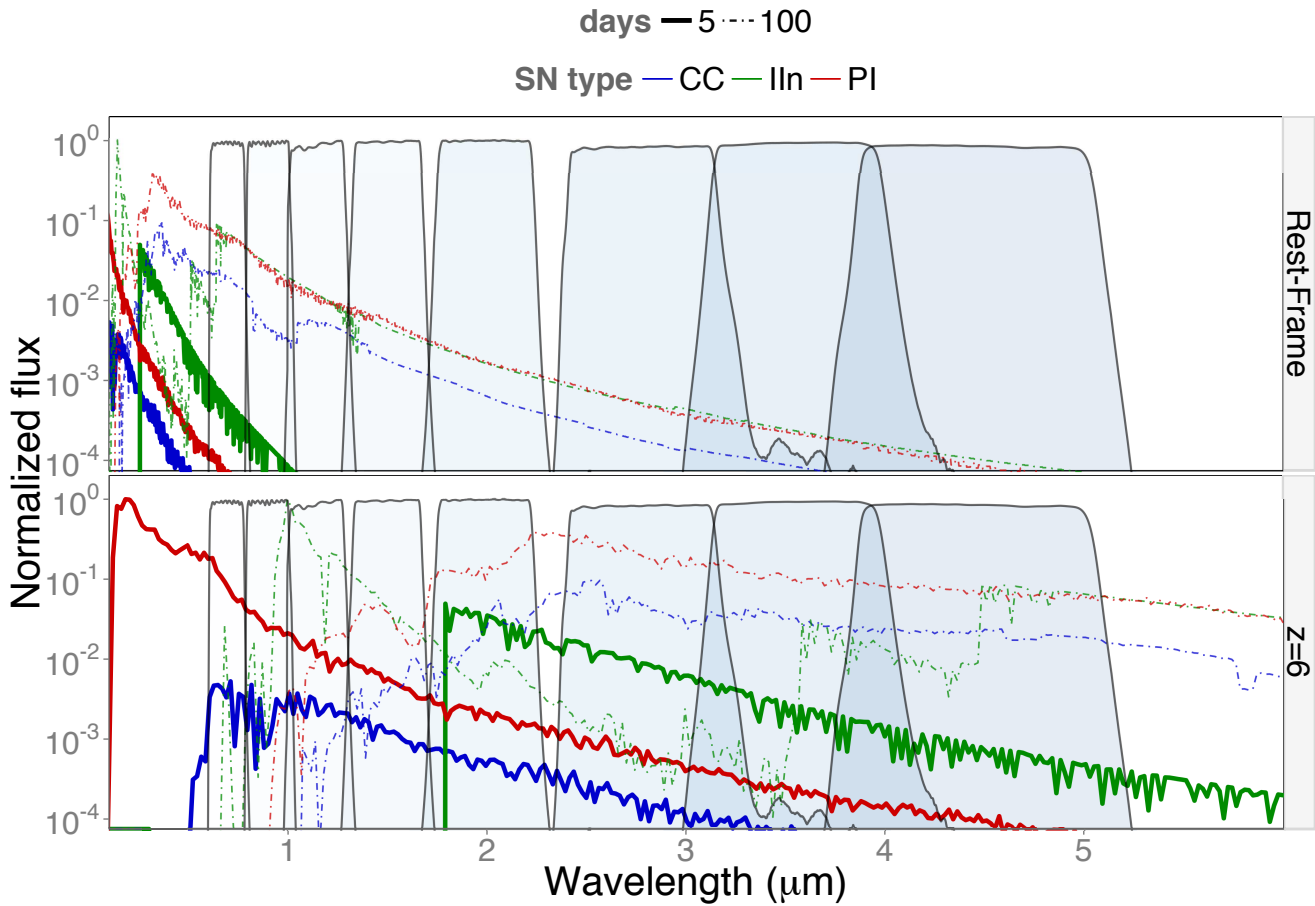}
\caption{ Spectral evolution of the z40G CC SN (blue lines), the a02 Type IIn (green lines), and the u200  PI SN (red lines) in the rest-frame (upper panel) and at \emph{z} = 6 (lower panel).   Fireball spectra are at $\approx$ 5 days 
(solid  thick lines), and 100 days (dashed thin  lines).  In the background we show the 8 NIRCam filters: F070W, F090W, F115W, F150W, F200W, F277W, F356W, F444W from left to right.} 
\label{fig:SNeFilter}
\end{figure*}

\begin{figure*}
\centering
\includegraphics[scale=0.35]{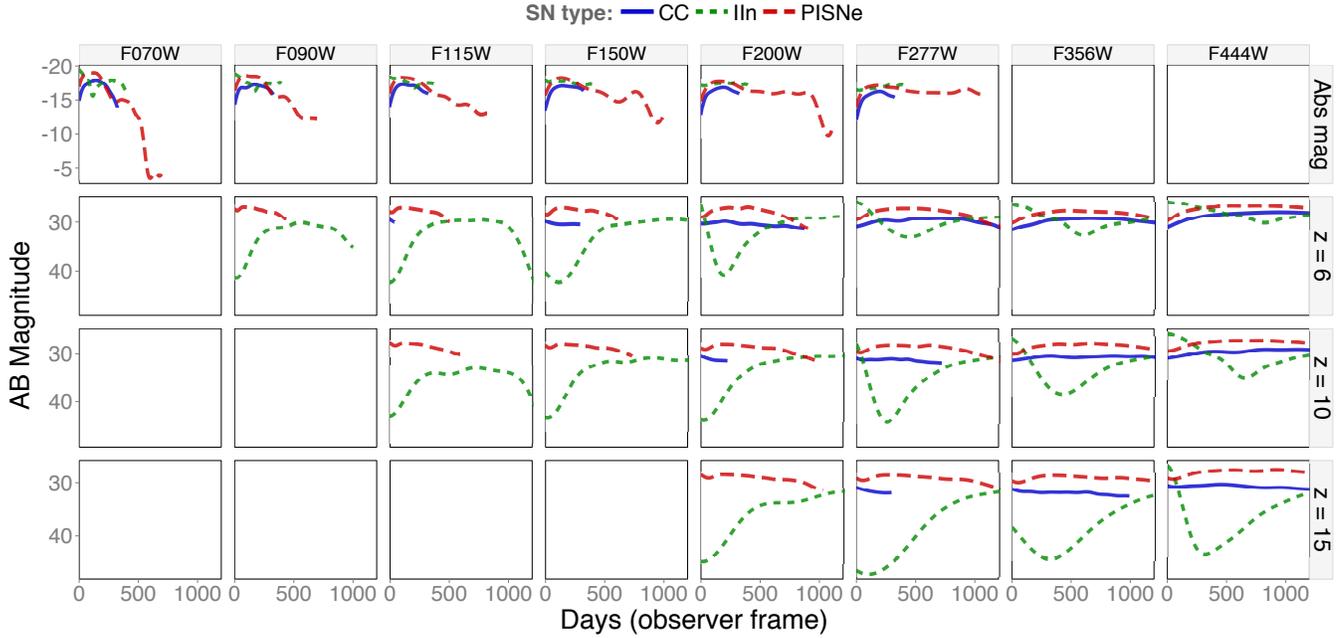}
\caption{Light-curve evolution of the  z40G CC SN (blue solid line), the a02  Type IIn (green dashed line), and the u200 PI SN (red long-dashed line).  LCs are at  10 pc,  $z$ =  6, 10 and 15. }
\label{fig:LCs_perf}
\end{figure*}

In Fig. \ref{fig:SNezoo} we show an example of CC, Type IIn, and PI SN spectral evolution from breakout to 100 days in the rest frame. The evolution of the PI and CC SN 
SEDs over time is mainly governed by two processes: (i) the fireball expands and cools, and its spectral cutoff advances to longer wavelengths; (ii) the ejecta and envelope, 
which were ionized by the breakout radiation pulse, begin to recombine and absorb photons at the high energy end of the spectrum, as evidenced by the flux that is blanketed 
by lines at the short-wavelength limit of the spectrum.  At later times, flux at longer wavelengths slowly rises due to the expansion of the surface area of the photosphere.  The 
SEDs are blue at earlier times and become redder at later times as the expanding blast cools.  In the Type IIn explosion, the shock driven by the ejecta through the shell 
causes the jump in luminosity after $\sim 30$ days.  The magnitude of the jump depends on the density of the shell, with diffuse shells allowing more radiation to pass through 
them.  A comparison between the  SED evolution for different SN types in the rest-frame and at redshift 6 are shown in Fig. \ref{fig:SNeFilter}. \footnote{Note that we are zooming the SED, and few features might  be missing in current  plot,  such as P-Cygni profiles in CC SNe. We refer to the original articles \citep{Whalen2013a,Whalen2013b,Whalen2013c,Whalen2013d} for a deep discussion of physical properties of each particular SN.}
For higher redshifts the overall SED is shifted to redder bands allowing a first estimation of the redshift of the source by dropout techniques (Fig. \ref{fig:LCs_perf}). 

In Fig. \ref{fig:LCs_perf}  we show LCs in the rest frame and at redshifts 6,
10 and 15 for comparison. The CC SNe exhibit plateaus in
the optical bands in the rest frame.  But high-redshift Type
IIP (CC) SNe do not exhibit plateaus in the NIR because emission
in the much bluer bands of origin in the rest frame rapidly 
declines due to the onset of line opacity at early times in the 
fireball (compare Fig. 4 with Fig. 7
from \citet{Kasen2009}, in which the U band flux falls off rapidly in comparison to the 
plateaus in the I, R and V band fluxes  \citep[see also][]{pan2012}).  The absence of a plateau in the bluer bands in the rest
frame of high-\emph{z} CC SNe is key to their detection in the NIR today. 
Otherwise, they might not be recognizable as transients.

%------------------------------------------------------------------------------------------------------------------------------------------------------------------------------------------------------------------------%

%------------------------------------------------------------------------------------------------------------------------------------------------------------------------------------------------------------------------%

%------------------------------------------------------------------------------------------------------------------------------------------------------------------------------------------------------------------------%

%------------------------------------------------------------------------------------------------------------------------------------------------------------------------------------------------------------------------%

\begin{table*}\centering
\label{tab:whalen_model}
\tikzset{ 
    table/.style={
        matrix of nodes,
        row sep=-\pgflinewidth,
        column sep=-\pgflinewidth,
        nodes={
            rectangle,
            draw=black,
            align=center
        },
        minimum height=1.5em,
        text depth=0.8ex,
        text height=2.2ex,
        nodes in empty cells,
        every even row/.style={
            nodes={fill=gray!20}
        },
        column 1/.style={
            nodes={text width=2.8em,font=\bfseries}
        },
         column 6/.style={
            nodes={text width=2.8em,font=\bfseries}
        },
        row 1/.style={
            nodes={
                fill=gray!40,
                text=black,
                font=\bfseries
            }
        }
    }
}

\begin{tikzpicture}

\centering
\matrix (first) [table,text width=6em]
{
Model  & SN Type &$M_{\star} (M_{\bigodot})$  & $E (10^{51}\rm  erg)$ & $Z (Z_{\bigodot})$& Model  & SN Type &$M_{\star} (M_{\bigodot})$  & $E (10^{51}\rm  erg)$ & $Z (Z_{\bigodot})$\\
u15B   &   CC  &  15  &  0.6  &  $10^{-4}$  &  u150  &  PI   &          150            &   9.0   &  $10^{-4}$    \\
u25B   &   CC  &  25  &  0.6  &  $10^{-4}$  &  u175  &  PI   &          175            &  21.3  &  $10^{-4}$    \\
u40B   &   CC  &  40  &  0.6  &  $10^{-4}$  &  u200  &  PI   &          200            &   33    &  $10^{-4}$    \\
u15D   &   CC  &  15  &  1.2  &  $10^{-4}$  &  u225  &  PI   &          225            &  46.7  &  $10^{-4}$    \\
u25D   &   CC  &  25  &  1.2  &  $10^{-4}$  &  u250  &  PI   &          250            &  69.2  &  $10^{-4}$    \\
u40D   &   CC  &  40  &  1.2  &  $10^{-4}$  &  z175  &  PI   &          175            &  14.6  &        0            \\
u15G   &   CC  &  15  &  2.4  &  $10^{-4}$  &  z200  &  PI   &          200            &  27.8  &        0            \\
u25G   &   CC  &  25  &  2.4  &  $10^{-4}$  &  z225  &  PI   &          225            &  42.5  &        0            \\
u40G   &   CC  &  40  &  2.4  &  $10^{-4}$  &  z250  &  PI   &          250            &  63.2  &        0            \\
z15B    &  CC  &  15   &  0.6  &        0         &   a00   &   IIn     &  $M_{sh}=0.1$    &   2.4   &        0            \\
z25B    &  CC  &  25   &  0.6  &        0         &   a01   &   IIn     &  $M_{sh}=1.0$    &   2.4   &        0            \\
z40B    &  CC  &  40   &  0.6  &        0         &   a02   &   IIn     &  $M_{sh}=6.0$    &   2.4   &        0            \\
z15D    &  CC  &  15   &  1.2  &        0         &   a03   &   IIn     &  $M_{sh}=10$     &   2.4   &        0            \\
z25D    &  CC  &  25   &  1.2  &        0         &   a04   &   IIn     &  $M_{sh}=20$     &   2.4   &        0            \\
z40D    &  CC  &  40   &  1.2  &        0         &  z15G &   CC    &          15               &  2.4    &        0            \\
z25G    &  CC  &  25   &  2.4  &        0         &  z40G &   CC    &          40               &  2.4    &        0           \\
};

\end{tikzpicture}
\caption{Properties of CC \citep{Whalen2013c}, IIn \citep{Whalen2013d} and PI SNe \citep{Whalen2013a,Whalen2013b} explosions: progenitor mass ($M_{\star}$), SN energy (E), and 
metallicity ($Z$).  The Type IIn explosions all have a progenitor mass of 40 $M_{\bigodot}$, but different shell masses and thus luminosities.}
\end{table*}
\section{Synthetic  observations}
\label{sec:survey}

%------------------------------------------------------------------------------------------------------------------------------------------------------------------------------------------------------------------------%

We generate synthetic LCs with a modified version of the \textit{SuperNova ANAlysis} \citep[\textsc{snana}; ][]{Kessler2009b} LC simulator, as in DS13.   It is a complete 
package for SN LC analysis, and accepts astrophysical characteristics of the source, IGM (see appendix  \ref{sec:IGM}), and survey parameters as inputs.  The SNe are 
specified by their source frame SEDs and $\dot{n}_{\rm SN}$.  The models simulated here span a wide range of progenitor parameters to emulate the diversity of SEDs 
expected in a real survey (Table 1).  The probability of each SN is determined by the given IMF as described in Section \ref{sec:SFH}.  Other physical elements like IGM 
filtering and the primary reference star, which defines the magnitude system to be used, are convolved with the specific filter transmissions through the construction of 
\textit{k}-correction tables.  Such tables transform the fluxes from source frame to the observer frame, on top of which extinction by the Milky-Way is applied using full-sky 
dust maps \citep{Schlegel1998}\footnote{Our observation fields were  centered around Hubble Ultra Deep Field \citep{beckwith2006}.}.  We also include telescope specifications such as CCD characteristics, field of view (FOV), point spread function (PSF) and pixel scale 
(Table \ref{tab:t_I}).  AB magnitudes are used for all the simulations presented here.    

The instrument characteristics in our synthetic observations are based on NIRCam \citep[appendix \ref{sec:JWST}; see also][]{Gardner2006}.  All simulations assume individual 
integrations of $10^2$ s, which can sometimes be co-added to create longer exposures.\footnote{Note that $\sim 10^3$ s is the limit for individual exposures due to cosmic 
ray contamination in the line of sight \citep{Gardner2006}.}  The two strategies tested here are the same as those in DS13 (Table 2):  

\paragraph* {Strategy 1 (str1)}
One pointing per year for each of the six reddest filters in NIRCam (F115W-F444W), to maximize our ability to identify high-redshift sources from 
non-detections in the bluest filters (dropouts).

\paragraph* {Strategy 2 (str2)}
Three pointings per year separated by two-month gaps in filters F150W and F444W, to maximize the number of SN detections.

\vspace{0.2in}

Both strategies assume a 5 yr survey time, with $\sim$ 730 hours of telescope time per year and a sky coverage of 0.06 per cent.\footnote{NIRCam consists of two fully redundant optical trails, which allows simultaneous observations of the same FOV in 2 filters, one in lower (F070W-F200W) and the other in higher (F277W-F444W) wavelengths \citep{Gardner2006}.  Taking advantage of this feature, our strategies were built so that each pointing gathers observations in 2 filters through 100s.} Finding at least a few detections near the LC 
peak is crucial to our goal of identifying these events as transients and photometrically categorizing them by SN type \citep[e.g., ][]{ishida2013}.  Therefore, for an event to 
be labelled a SN we require its detection at a minimum of one epoch before and after maximum brightness.  We also require its detection at three epochs in at least one filter 
above the background limit and selection cuts on signal to noise (S/N) ratio $\geq$ 2 (cut1) and $\geq$ 3 (cut2) in at least one filter. 

\begin{figure*}
\centering

\includegraphics[scale=0.25]{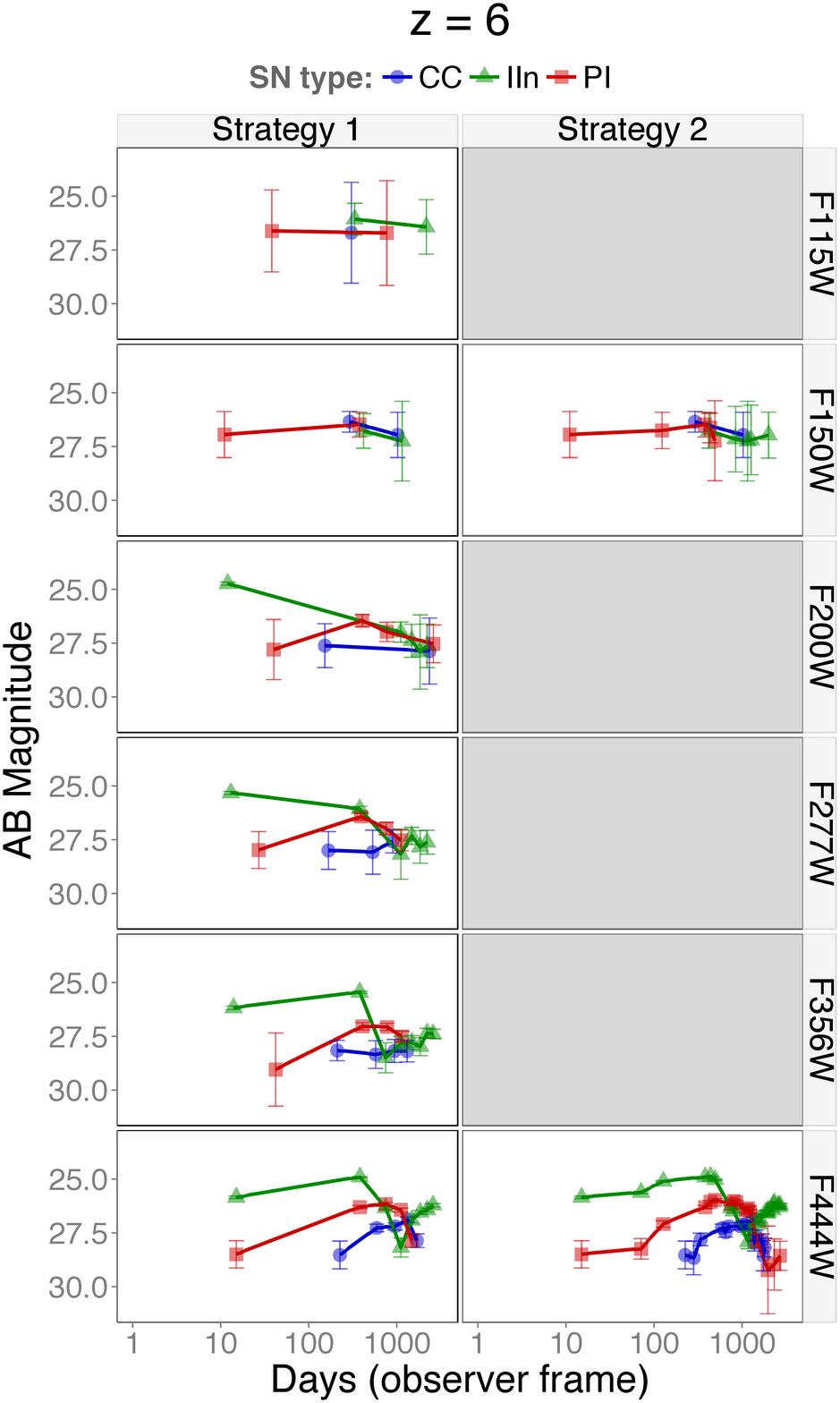}
\hspace{0.15 cm}
\includegraphics[scale=0.25]{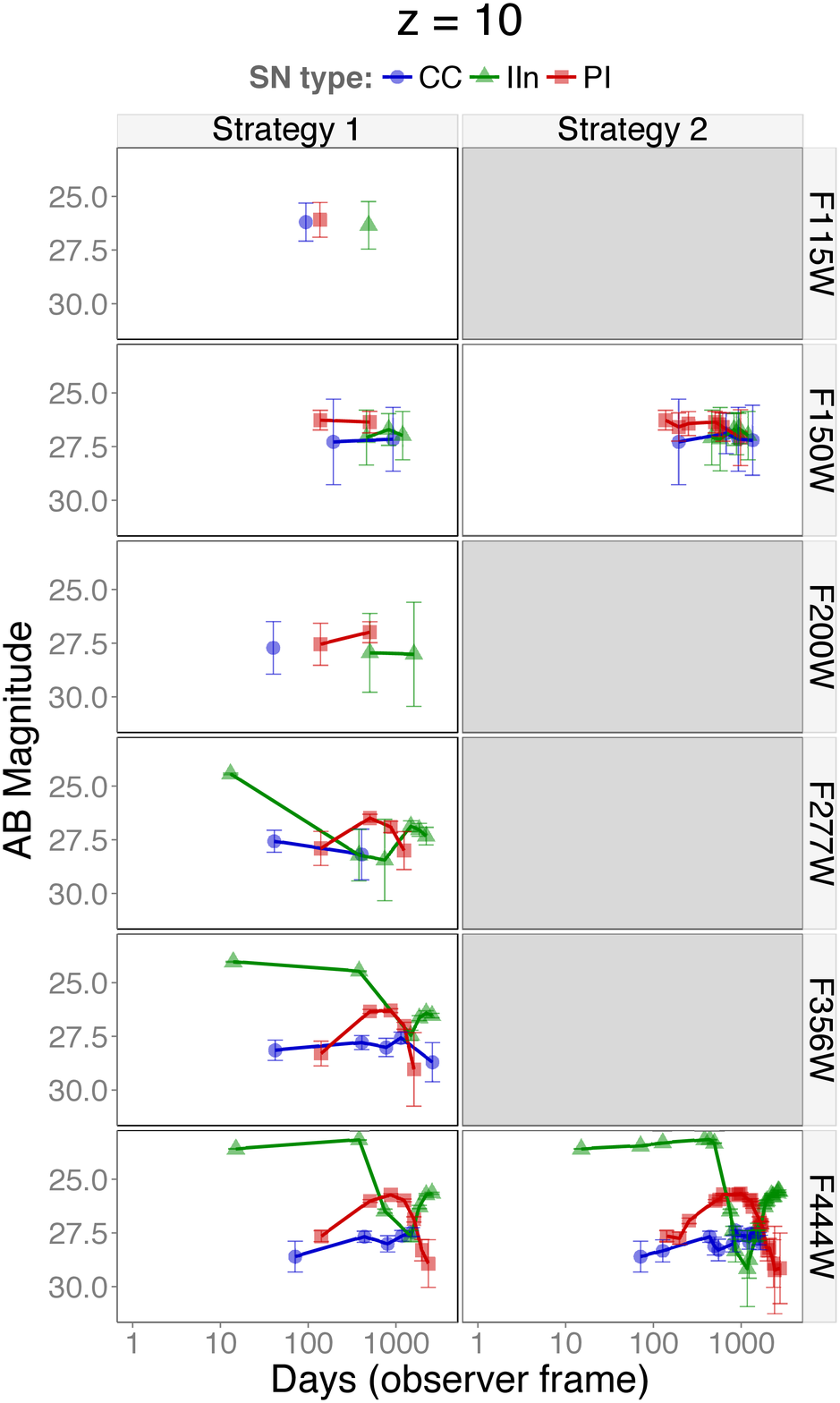}
\hspace{0.15 cm}
\includegraphics[scale=0.25]{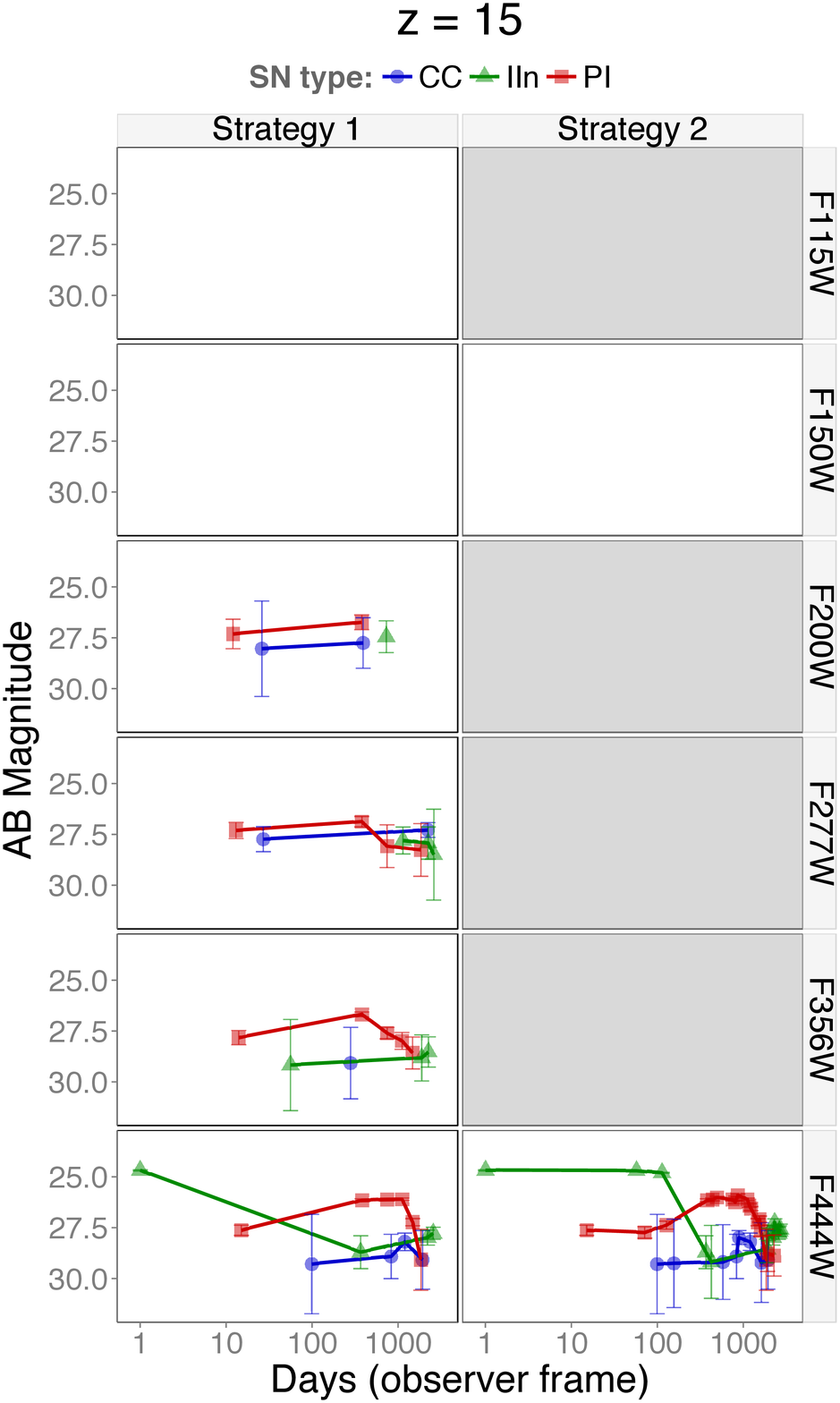}
\caption{SEDs of CC, Type IIn and PI SNe as they would be observed by the \textit{JWST} NIRCam in our two observational strategies. We show each LC at z = 6 (left panel), 
10 (central panel) and 15 (right panel).  The y-axis is the observed magnitude and the x-axis is the time in the observer frame.  The high-redshift events vanish from the 
bluest filters and their LCs last longer due to time dilation. The grey panels represent the filters not used by a specific strategy, and in the blank panels there were no detections.}

\label{fig:LCs}
\end{figure*}

In  Fig. \ref{fig:LCs}, we show examples of LCs obtained by both strategies. The error bars denote the uncertainties in the magnitude measurements due to the background 
on the sky, photon statistics and instrument calibration. The errors are summed in quadrature and considered over an effective aperture based on PSF fitting, which we take 
to be Gaussian (see appendix \ref{sec:noise}). From this figure, we conclude that detecting a SN in the six reddest filters indicates an event at moderate redshift, $z \lesssim 
10$, while a detection in only the four reddest filters suggests a SN at $z\gtrsim 15$. 

\begin{figure*}
\centering
\includegraphics[scale=0.35]{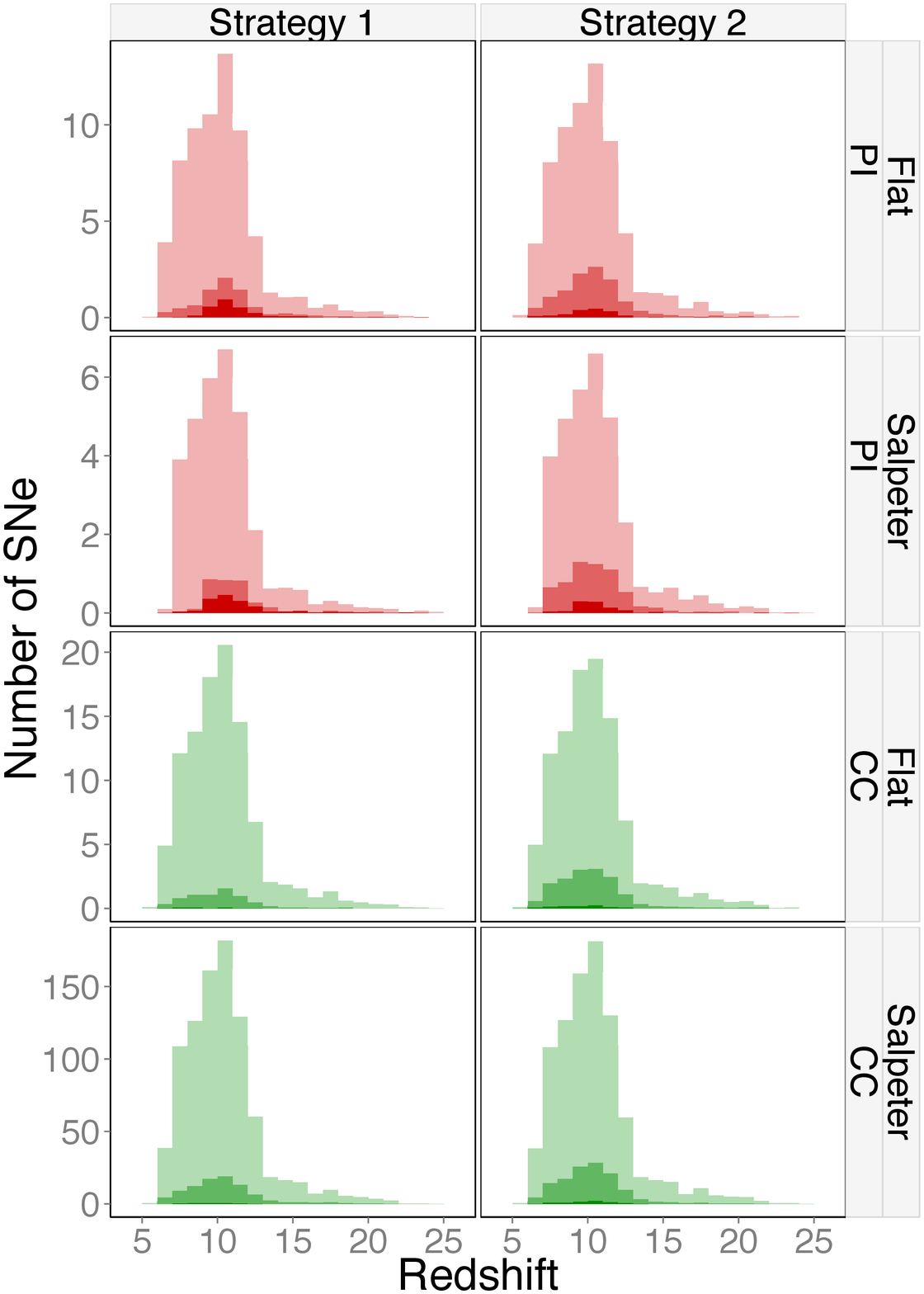}
\hspace{1.5 cm}
\includegraphics[scale=0.35]{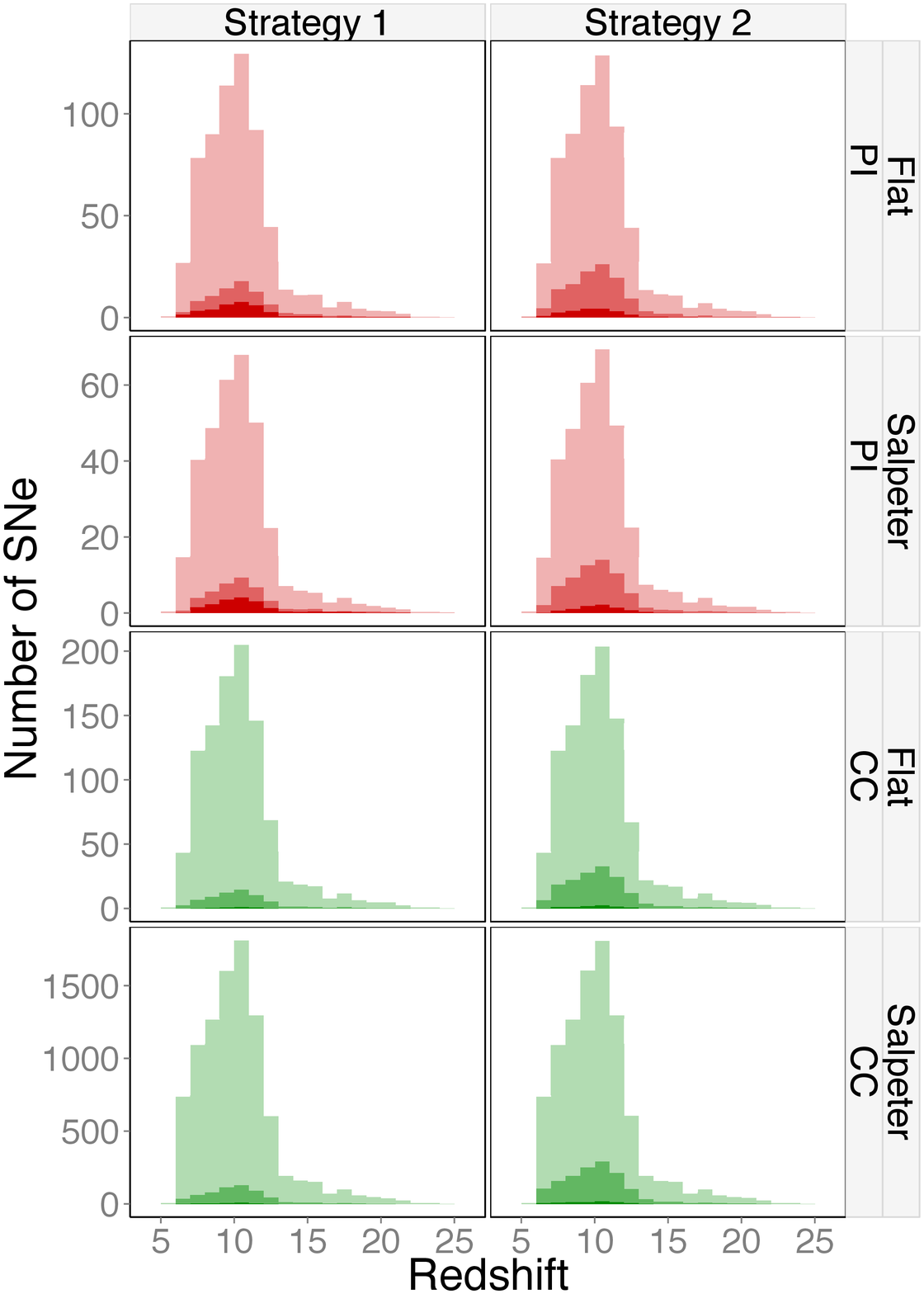}
\caption{Number of total SN explosions during 5 yr mission  in the observed field (lightest colours),  and number of detected SNe after cut1 (middle colours) and cut2 (darkest colours) as a function 
of redshift for each model (CC, Type IIn and PI SNe), averaged over 200 realizations.  Left panel:  SFR1; right panel:  SFR10.  The left and right columns in each panel are for
our two survey strategies, str1 and str2, and the rows indicate SN model and IMF. Note that not all axes are scaled the same.
}
\label{fig:histogram}
\end{figure*}

Fig. \ref{fig:histogram} shows the number of detections in comparison to the total number of SNe in the field.  A few important conclusions can be drawn by inspecting the 
histograms:
 
\begin{itemize}

\item str2 clearly optimizes the detection rate of all SN types for cut1 for both SFR1 and SFR10 (using the same amount of time), but at the cost of getting information in a 
smaller number of filters.  However,  str1 seems to be more effective at detecting the higher quality LCs imposed by cut2. This is mainly due to the larger number of filters 
in this strategy, which increases the likelihood of finding a brighter event in at least one filter (the exact number of detections are presented later in Section 5 and Appendix 
B). 

\item The relative number of SNe detected of each type is strongly correlated with the underlying IMF, and therefore might be used as a probe of the Pop III IMF, as 
discussed later in this article.  For instance, CC detections increase by almost an order of magnitude from the flat to the Salpeter IMF because the mass distribution of 
the latter is biased toward less massive stars.  

\item Given a fixed amount of telescope time, changing the observation strategy might increase the number of detections.
\end{itemize}

To be more quantitative, for the most optimistic case (SFR10+cut2) we obtain $39 \pm 9$ ($67 \pm13 $) PI SNe, and $535 \pm 33$ ($1451 \pm70 $) CC SNe with str1 
(str2) for the Salpeter IMF.  We  find  $76 \pm 12$ ($124\pm 17$) PI SNe and $61\pm 10$ ($159 \pm 22$) CC SNe with str1 (str2) for the flat IMF.  For Type IIn SNe, 
the detection rate is negligible, with $\sim 1$ event every 10 yr in the most optimistic case, so we do not bin them here.\footnote{All numbers are an average over 200  
simulations.}  This is a clear indication that \textit{JWST} can observe a large number of  Pop III SNe, particularly CC SNe.  Once potential candidates are detected, the 
brightest sources will be suitable for more detailed spectroscopic followup, which can place strong limits on their metal content.\footnote{For low-metallicity objects, the 
ratio of oxygen lines to Balmer lines, such as  [OIII]/H$\beta$, provides a linear measurement of metallicity.}  

Whilst the PI SNe  are brighter enough to be detected by \textit{JWST}  at redshifts as high as $z\sim 25$, the scarcity of such  
events makes their detection at the lowest redshifts more tangible. Our results  are consistent with \citet[][Fig. 8]{Hummel2012} for an equivalent amount of time and FOV, although, some precaution must be taken when comparing 
numerical results. While they used a more stringent $10-\sigma$ detection limit, they did not consider multiple observations in different filters, which demands a considerably amount of time.  Nonetheless, similar to us, they conclude that
the strategy more likely to succeed in detecting primordial SNe must rely on  
multiple field search strategy. 
 
%------------------------------------------------------------------------------------------------------------------------------------------------------------------------------------------------------------------------%

\begin{table*}
\caption{Series of observational search strategies and selection cuts. }
\begin{center}
\begin{tabular}{lccccc}
\hline 

 Run   & Sky  (per cent) & Cadence & NIRCam Filters & \textit{JWST} time per yr  \\
 
 \hline

 Strategy 1    &{0.06}      &{One pointing per year}    & F115W-F444W     & {730 h}  \\
 Strategy 2    & {0.06}   & {Three pointings per year, two-month gaps} & {F150W, F444W} &{730 h} \\
\\
 \hline
  \multicolumn{5}{c}{Detectability requirements} \\
  \hline
\multirow{2}{*}{Detection} &\multicolumn{3}{c}{At least one epoch before and one epoch after maximum}&\\
 			&\multicolumn{3}{c}{At least three epochs in one filter}&\\
\\			
 		
 		Minimum data quality & \multicolumn{3}{c}{cut1: at least 1 filter with S/N$>$2}&\\
		& \multicolumn{3}{c}{cut2: at least 1 filter with S/N$>$3}&\\
 \hline
\end{tabular}
\end{center}
\label{tab:strategy}
\end{table*}

%------------------------------------------------------------------------------------------------------------------------------------------------------------------------------------------------------------------------%

%------------------------------------------------------------------------------------------------------------------------------------------------------------------------------------------------------------------------%

%------------------------------------------------------------------------------------------------------------------------------------------------------------------------------------------------------------------------%

\section{Probing the Primordial IMF }
\label{sec:IMF}

The ultimate goal of a complete theory of primordial star formation is to predict the Pop III IMF from first principles \citep{Bromm2013}.  Finding the first cosmic explosions      
would be a leap forward in our understanding of the basic processes behind the formation of the first stars.  Since each SN type to some degree encodes the mass of its  
progenitors, their relative rates can place firm constraints on the Pop III IMF.      

But inferring the nature of a progenitor from its SN may not be simple. Even the progenitors of well-studied Type Ia SNe are not fully understood \citep{Hillebrandt2000,
Wang2012}, although it is generally believed that they fall into one of two types:  single-degenerate and double-degenerate mergers \citep{Maoz2013}.  Techniques like 
integral field spectroscopy have been used to obtain both spatial and spectral information on the explosion site, allowing the identification of the parent population of the 
star.  While many stellar evolution models have been proposed to characterize the progenitor stars, these predictions are still poorly supported by observations \citep{
Kuncarayakti2013}.  Given the aforementioned sources of uncertainties, instead of reconstructing the Pop III IMF directly from the data, we seek the answer to a more 
straightforward question:  is it possible to distinguish a Salpeter IMF from a flat IMF by counting the relative observed rates of CC and PI SNe?  In other words, we want to 
test the following null ($\mathcal{H}_0$) and alternative hypotheses ($\mathcal{H}_a$):
\begin{description}

\item $\mathcal{H}_0$: The relative observed frequencies of CC and PI SNe are independent of the IMF.

\item

\item  $\mathcal{H}_a$: The relative observed frequencies of CC and PI SNe are not independent of the IMF. 

\end{description}
The CC and PI SNe can be treated as categorical variables,\footnote{Categorical data is composed of variables that can be separated into mutually exclusive classes, but 
there is no intrinsic ordering to the categories.} and their degree of dependence on the IMF can be investigated using the formalism of contingency tables \citep{
Becker1996}.  A $2\times2$ contingency table is a way of visualizing  the dependence between two categorical variables (columns) and a given factor (rows).  Examples 
of such tables for three of the scenarios in Fig. \ref{fig:histogram} are shown in Tables \ref{tab:flat}, \ref{tab:SFR10} and \ref{tab:all}. To measure the degree of correlation 
between them, we use the $\chi^2$ test of independence.  

Let $O_{ij}$ be the original contingency table and $E_{ij}$ be the one we would expect to measure if the categorical variables were independent. The latter is constructed   
by multiplying the sum of rows ($R_{i}$) and columns ($C_{j}$) of $O_{ij}$ and then dividing by the total number of observations ($\mathcal{N}$), $E_{ij} = R_{i}\times C_
{j}/\mathcal{N}$.  In the tables described above, the $E_{ij}$ for each case is shown in parentheses.  We reject $\mathcal{H}_0$ if the $p$-value\footnote{$p$-value is the 
probability of obtaining a test statistic at least as extreme as the one that was actually observed, assuming that $\mathcal{H}_0$ is true \citep[e.g.,][]{goodman2008}.} of 
the following $\chi^2$ test is below a given significance level $\alpha$,  
\begin{equation}
\chi^2 = \sum_{i,j}^{2} \frac{(O_{ij}-E_{ij})^2}{E_{ij}}, 
\label{eq:chi}
\end{equation}
where typical values for alpha are 0.01 or 0.05, representing 99 and 95 per cent confidence levels, respectively.  For instance, the $\chi^2$ test for Table \ref{tab:flat} is given 
by 
\begin{eqnarray}
\chi^2 &=&  \frac{(339-449.81)^2}{449.81}+\frac{(9312-9201.19)^2}{9201.19}\\
&+&\frac{(124-13.18995)^2}{13.18995}+\frac{(159-269.81)^2}{269.81} \approx 1005. \nonumber
\end{eqnarray}
In order to emulate a realistic scenario, we compared each of the synthetic observations with a basis model.  Thus, the first row of all the tables are equal (representing 
our basis model) while the second row represents the observed sample for each case. Our fiducial model is composed of the total number of SNe expected to appear in 
the {\it JWST} survey area, assuming a Salpeter IMF and SFR10. 
 
Tables \ref{tab:flat} and \ref{tab:salp} show the contingency tables for observed samples with flat and Salpeter IMFs, respectively.  Both tables assume SFR10 and cut1 
(for the others, see appendix \ref{sec:tables}).  An observed sample whose IMF is flat is easily seen not to be Salpeter with a 99.95 per cent  confidence level because $\chi^2 
\approx 1005$, with a $p$-value $\lesssim$ 0.0005.  Furthermore, the observed sample whose IMF is Salpeter is not wrongly rejected due to observational bias, with $
\chi^2 = 2.79$ and a $p$-value = 0.095.  From Tables B1-B3, we see that the flat IMF is systematically rejected in all scenarios, showing high values of $\chi^2$: 365.5
(a $p$-value $\lesssim$ 0.0005 for str2+cut2, 893.3 (a $p$-value $\lesssim$ 0.0005) for str1+cut1, and 755.6 (a $p$-value  $\lesssim$ 0.0005) for str1+cut2. 

In Tables B4-B6, we show the comparison with an observed sample generated from a Salpeter IMF.  All values of $\chi^2$ are lower than for the flat IMF,  indicating a 
better agreement with $\mathcal{H}_0$. Thus, we can already conclude that the $\chi^2$ analysis is able to indicate the most likely IMF between two scenarios. 

 More
insights can be gained from the relative risk \citep[RR; ][]{Zhang1998}.  If the samples are drawn from the same IMF, RR approaches 1, while it diverges from 1 when 
the samples are from two divergent IMFs.  In other words, RR measures the probability of a certain event occurring in one group versus another.  
Therefore all numbers between brackets,  representing  the $E_{ij}$ table,  are constructed  in such a way that $RR  \approx 1$. Hence, our method  determines  if deviations from  RR = 1 are driven by random fluctuations or  by  some intrinsic effect of the underlying IMF.  
An interesting effect 
arising from the RR analysis is the Malmquist bias, which leads to the preferential detection of intrinsically brighter objects.  Comparing Tables 4 and B4, the RR for PI
SNe changes from 0.796  to 0.433 merely by changing the S/N requirement.  This means that the probability of observing a PI SN is almost twice as high when cut2 is 
applied instead of cut1. This happens because the more stringent the detection limit, the more biased our sample will be towards brighter objects.
      
While the $\chi^2$ analysis is useful for deciding which IMF better fits the data, it is also important for  
determining if
the observed and theoretical 
samples are drawn from the same IMF in a self-consistent manner.  For instance, because of the low number of events from str1, the $p$-value analysis  \textit{per se}  
wrongly rejects the Salpeter IMF, finding $p$-values $< 0.05$ (Tables B5 and B6).  In other words, even if the fiducial and observed models are drawn from the same IMF,   
poor statistics may cause a bias in our conclusion.  Thus, determining the proper size of a sample is a crucial step for experiment design, and it constitutes a branch in 
statistics known as power analysis that we discuss in greater detail below.  
  
%------------------------------------------------------------------------------------------------------------------------------------------------------------------------------------------------------------------------%
    
\subsection{Sample size}
\label{sec:power}

Once we are convinced the above procedure is satisfactory for our purposes, we still need to optimize the use of telescope resources in order to design a realistic survey.  
Therefore, it is important to determine the minimum sample size necessary to measure a desired effect or to estimate how much should we trust a particular result, given  
the sample size allowed by observational constraints.  Otherwise, even if the survey is performed perfectly and analyzed properly, wrong conclusions could still be made
due to the lack of statistical power ($\mathcal{P}$).  $\mathcal{P}$ can be understood to be the probability of rejecting $\mathcal{H}_0$ when $\mathcal{H}_a$ is true.  In 
our case, it represents the ability to detect true deviations from the Salpeter IMF in the observed sample.  Therefore, while $\alpha$ represents  the probability that 
$\mathcal{H}_0$ will be rejected when it is actually true (a wrong decision), $\mathcal{P}$ quantifies the likelihood of rejecting  $\mathcal{H}_0$ when it is actually false (a 
correct decision).

$\mathcal{P}$ depends on the effect size ($\mathcal{W}$), confidence level, $\alpha$,  and number of detections,  $\mathcal{N}$.  $\mathcal{W}$ determines the deviation from $\mathcal{H}_0$ that one would 
expect to detect.  The number increases with the degree of discrepancy between the distribution of $\mathcal{H}_a$ and $\mathcal{H}_0$, 
\begin{equation}
\mathcal{W} = \sqrt{\sum_{i=1}^2 \frac{(P_{0i}-P_{Ei})^2}{P_{Ei}}}, 
\end{equation}
where $P_{0i}$ and $P_{Ei}$ are the proportions (counts divided by total number of observations) in each cell for the  $O_{ij}$ and $E_{ij}$ contingency tables, respectively. 
It can be written in terms of $\chi^2$ as $\mathcal{W} = \sqrt{\chi^2/\mathcal{N}}$.  For our basis model shown in Table \ref{tab:fiducial}, $\mathcal{W} = 0.44$. Traditionally,  
$\mathcal{W} = 0.1, 0.3, 0.5$ are considered thresholds for small, medium and large differences, respectively \citep{Cohen88}. The $\chi^2$ statistics follows a central  $\chi^2_{\kappa}$ distribution when $\mathcal{H}_0$ is true
\\
\begin{equation}
\chi^2_\kappa (x) = \frac{x^{\kappa/2-1}e^{-x/2}}{\Gamma\left(\frac{1}{2}\kappa\right)2^{\kappa/2}},
\end{equation}
\\
where $\Gamma$ is the gamma function \citep[e.g.,][]{Abramowitz1965} and $\kappa$ stands for degrees of freedom (equal to one for $2\times2$ tables). 
 While it follows a noncentral $\chi
^2_{\kappa}(\lambda)$ distribution  if $\mathcal{H}_0$ is false, 
\\
\begin{equation}
\chi^2_\kappa (x,\lambda) = \frac{e^{-(x+\lambda)/2}x^{(\kappa-1)/2}\sqrt{\lambda}}{2(\lambda x)^{\kappa/4}}I_{\kappa/2-1}\left(\sqrt{\lambda x}\right), 
\end{equation}
\\
where $I_n (x)$ is the modified Bessel function of the first kind \citep[e.g.,][]{Abramowitz1965}.  $\mathcal{P}$ represents our ability to discriminate between $\chi^2_{\kappa}$ and $\chi
^2_{\kappa}(\lambda)$ at a certain  $\alpha$, 
\begin{equation}
\mathcal{P} = P(\chi^2_{\kappa}(x,\lambda) \geqslant \chi^2_{\kappa;1-\alpha} (x)), 
\end{equation}
where $\lambda = \mathcal{W}\mathcal{N}$ is the noncentrality parameter. If both samples are equal, $\mathcal{H}_0$ is true and $\mathcal{W} = \lambda = 0$. Hereafter, all calculations are performed with $\alpha = 0.05$ or at 99.5 per cent confidence level.  

\begin{figure}
\centering
\includegraphics[scale=0.42]{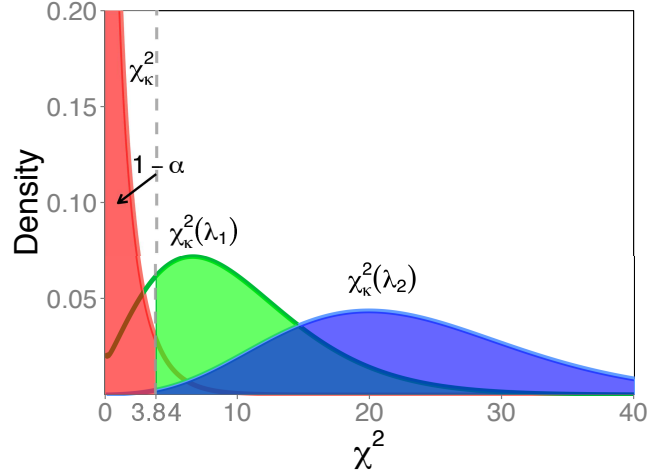}
\caption{The  red curve  denotes the $\chi^2_\kappa$ distribution under $\mathcal{H}_0$. The blue and green curves are 
 the distribution under $\mathcal{H}_a$ for noncentrality parameter $\lambda_1 = 20\times0.44$ (20 detections, $\mathcal{W}$ = 0.44 ) and $\lambda_2 = 50\times0.44$ (50 detections, $\mathcal{W}$ = 0.44) respectively.  The vertical dotted grey  line represents the $\chi^2$ value with a probability of 0.05 under
$\mathcal{H}_0$. If we get a $\chi^2$ statistic to the left of the line we conclude that it came
from the $\mathcal{H}_0$ and if it is to the right of the line then we conclude it came from
$\mathcal{H}_a$. The area under the alternative distribution to the right of the
line represents the power of the test. Thus, the higher  the number of
observations, more straightforward is to discriminate  $\mathcal{H}_a$  from $\mathcal{H}_0$. Which in our case represents the ability to conclude if the distribution is consistent with Salpeter of not. }
\label{fig:power_graph}
\end{figure}

In Fig. \ref{fig:power_graph} we show a graphical representation  of  $\mathcal{P}$. The  shaded red  area on the left of the grey-dashed line at $\chi^2 = 3.84$  represents  99.5 per cent confidence level for the $\chi^2_{\kappa}$.
In this context,    $\mathcal{P}$ is the area under  $\chi^2_{\kappa}(\lambda)$  falling on the right side of the grey-dashed line.   Therefore the rightmost  the 
 $\chi^2_{\kappa}(\lambda)$ distribution for  
 $\mathcal{H}_a$ is, the easier it will be to discriminate it  from the  $\chi^2_{\kappa}$ distribution.  
Higher values of $\lambda$ 
implies  
higher deviations from $\chi^2_{\kappa}$. Hence,  $\mathcal{P}$ grows in the same way as the number of detections. 

\begin{figure}
\centering
\includegraphics[scale=0.47]{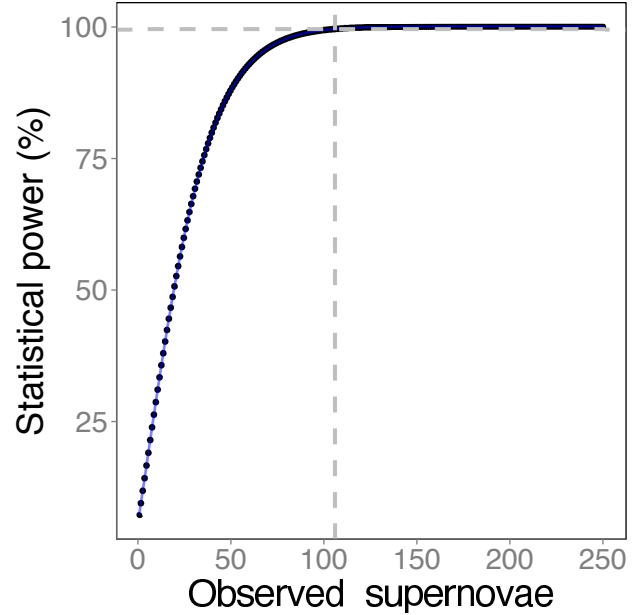}
\caption{Power estimation as a function of the number of detected SN, demonstrating that to achieve a power of 99.5\% (horizontal dashed gray line), it is necessary to 
detect at least 105 SNe (vertical dashed gray line).}
\label{fig:power}
\end{figure}

In Fig. \ref{fig:power} we show that $\sim 10^2$ SN detections are necessary to distinguish between a Salpeter and a flat IMF with the $\chi^2$ test.  It is important to note 
that the former number represents the number of SNe properly classified within their respective categories, not the total number of SNe appearing in the FOV.  This is a 
remarkable result because it shows that the Pop III IMF could be unveiled in the near future given a reasonable amount of telescope time. 

Finally, the  methodology suggested here is broad enough to  determine  which of  two   IMF  better describe the data.  We use the   Salpeter and  flat IMF as a benchmark of the test since these are two physically motivated models. Furthermore  our methodology defines a straightforward   way to determine the minimum sample  requirements to achieve a reliable result.

 %------------------------------------------------------------------------------------------------------------------------------------------------------------------------------------------------------------------------%

%--------------------------------------------------------------------------------------------------------------------------------------------------------------------------------------------------------------------------------%

%------------------------------------------------------------------------------------------------------------------------------------------------------------------------------------------------------------------------%

 \begin{table}

 \caption{Contingency table for the fiducial model and the case with flat IMF, SFR10, cut1 and str2.}
 \begin{center}
\begin{tabular}{lcc|cc}
\hline
 &  \multicolumn{2}{c}{SN type}\\
 \hline
  IMF & PI SN & CC & Total by IMF\\
  \hline
  $\rm Salpeter_{(all)}$ & 339 (449.81) & 9312 (9201.19)  & 9651\\
                                    
  $\rm Flat_{(cut1;str2)}$ &   124 (13.19) &159 (269.81) &283\\
\hline  
  Total by type & 463 & 9471 & 9934\\
  \hline
    \end{tabular}

  \begin{itemize}

 \item Relative risk: PI SN = $\frac{339/9651}{124/283}  \approx 0.080$\quad CC = $\frac{9312/9651}{159/283}  \approx 1.717$
\item $\chi^2 =  1005.068$ \quad  p-value $< 0.0005$
\end{itemize}
  \end{center}
   \label{tab:flat}

 \end{table}
%--------------------------------------------------------------------------------------------------------------------------------------------------------------------------------------------------------------------------------%  

%------------------------------------------------------------------------------------------------------------------------------------------------------------------------------------------------------------------------%
\begin{table}

 \caption{Contingency table for the fiducial model and the case with Salpeter IMF, SFR10, cut1 and str2.}
 \begin{center}
 \label{tab:SFR10}
\begin{tabular}{lcc|cc}
\hline
 &  \multicolumn{2}{c}{SN type}\\
 \hline
  IMF & PI SN & CC & Total by IMF\\
  \hline
  $\rm Salpeter_{(all)}$ & 339 (350.82) & 9312 (9300.18)  & 9651\\

  $\rm Salpeter_{(cut1;str2)}$ &   67 (55.18) &1451 (1462.82) &1518\\
\hline  
  Total by type & 406 & 10763 & 11169\\
 
  \hline
    \end{tabular}

  \begin{itemize}
   \item Relative risk: PI SN = $\frac{339/9651}{67/1518}  \approx 0.796$\quad CC = $\frac{9312/9651}{1451/1518}  \approx 1.009$
   \item $\chi^2 = 2.79$ \quad  p-value $= 0.095$

\end{itemize}
  \end{center}
   \label{tab:salp}

 \end{table}
%--------------------------------------------------------------------------------------------------------------------------------------------------------------------------------------------------------------------------------%

%--------------------------------------------------------------------------------------------------------------------------------------------------------------------------------------------------------------------------------%  
\begin{table}

 \caption{Contingency table for the total number of SNe expected to appear  in the  JWST survey area  for the fiducial model and the case with a flat IMF.}
 \begin{center}
 \label{tab:all}
\begin{tabular}{lcc|cc}
\hline
 &  \multicolumn{2}{c}{SN type}\\
 \hline
  IMF & PI SN & CC & Total by IMF\\
  \hline
  $\rm Salpeter_{(all)}$ & 339 (834.57) & 9312 (8816.43)  & 9651\\

  $\rm Flat_{(all)}$ &   638 (142.43) &1009 (1504.58) &1647\\
\hline  
  Total by type & 977 & 10321 & 11298\\
 
  \hline
    \end{tabular}

  \begin{itemize}
   \item Relative risk: PI SN = $\frac{339/9651}{638/1647}  \approx 0.091$\quad CC = $\frac{9312/9651}{1009/1647}  \approx 1.575$
   \item $\chi^2 = 2209.739$ \quad  p-value $< 0.0005$

\end{itemize}
  \end{center}
   \label{tab:fiducial}

 \end{table}
%--------------------------------------------------------------------------------------------------------------------------------------------------------------------------------------------------------------------------------%

\section{Supernova typing}
\label{sec:class}

The analysis delineated in the last section implicitly assumes that all  SNe have been  properly  classified into their  own groups.  Nevertheless,  once we have identified a  Pop III SNe, we must classify them into PISN/CC events and provide some estimate of the level of contamination. 
These are far from trivial  tasks  and a  detailed study of  all nuances  underlying the SNe photometric classification \citep{ishida2013}  demands a separate work. However, we   describe below how such problems can be tackled, demonstrating once more the potential of Pop III synthetic surveys as test grounds for future analysis.

Distinguishing between PI and CC SNe would be  fairly straightforward if one can  measure their redshift. In rest-frame, NIR  PISN light
curves last significantly longer than CC ones due to longer radiation diffusion timescales in their more massive
ejecta. Although, before performing a time consuming  spectroscopic follow-up on such faint targets, it is wise to develop an optimized strategy for photometric classification to identify potential candidates. This  subject is already highly developed within the type Ia SN community \citep{Kessler2010}, and the existing approaches are usually divided in two classes: empirical \citep[e.g., ][]{Newling2011,Richards2011,karpenka2012,ishida2013}   and template fitting \citep[e.g., ][]{Sullivan2006,Poznanski2007,kuznetsova2007,Kunz2007,Sako2008,Rodney2009,Sako2011} methods. The former uses magnitudes and/or fluxes  of a spectroscopically measured sample for training the method, which is then applied to the photometric sample. The latter, try to find spectral template and redshift which best fit the photometric observations using a library of well know observational or synthetic spectra.

\begin{figure}
\centering
\includegraphics[scale=0.47]{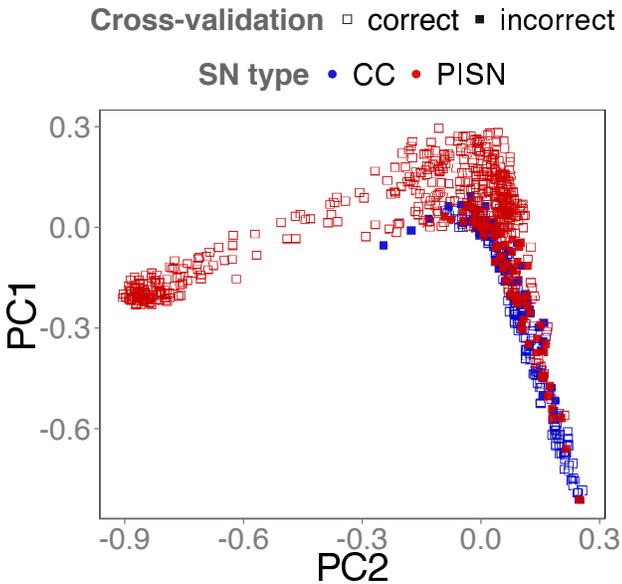}
\caption{Kernel Principal Components projection of the LCs from CC (blue symbols) and PI (red symbols).  Empty square represents SN correctly classified and filled square SN wrongly  classified  by the cross-validation using $k$-nearest neighborhood algorithm.}  
\label{fig:proj}
\end{figure}

Since we have not yet observed any Pop III stars,   which could in principle be used as a training sample,  the best  we can do is to base our analysis on synthetic LCs in order to estimate our potential power in  discriminating between  CC and  PI SNe. Following the procedure suggested by  \citet[][see appendix \ref{sec:kpca}]{ishida2013}, we present in Fig. \ref{fig:proj} the projections of CC and PISN LCs into a 2-dimensional space obtained from Kernel Principal Component Analysis (KPCA). Each point represents a LC observed in filters F200W, F277W, F356W and F444W with a cadence of two months in each filter. The dispersion of the projected  points are mainly  caused by the intrinsic variance of different SN models, S/N and redshift.  Each SN type in the figure is identified by colour, while the marker style indicates if an   object was correctly classified or not by a cross-validation procedure using the  $k$-nearest neighbor predictor within a \textit{Leave One Out} (LOO) cross-validation procedure (appendix \ref{ap:CV}).

In the context of this work, each iteration of the LOO algorithm can be interpreted as an attempt to classify a new unknown object given the synthetic   data set. Consequently, the cross-validation can be understood as an indicative of the  power of the method in discriminating between each class of SN. Nevertheless a more realistic prediction  requires the  application of the trained  classifier to a new set of  LCs whose classes are unknown.  Furthermore,  we should carry an exhaustive   study of the observational cadency, data quality, expected amount of spectroscopic confirmed and photometric objects of  each type, LC fitting method, number of used photometric bands,  and so on \citep{ishida2013}.  Such analysis is out of the scope of the present work, but planned to be carried in the future.

The confusion matrix for the objects present in  Fig. \ref{fig:proj} is given in Table \ref{tab:Confusion},  representing the number of SN correctly (diagonal)  and incorrectly (off-diagonal) classified during the cross-validation. For this example case, the method achieve an accuracy of $\sim$ 76.3  per cent. Assuming that the real SN will not be very different from our simulations, the results suggests an approximate
  upper limit for our ability to correctly discriminate between these objects.  It has direct implications in the analysis  developed  in Section \ref{sec:power}, which  implicitly assumes an error-free classification of the LCs.

In other words,  standard power analysis  does not take into account the reliability ($\mathcal{R}$) of the measurement.  According to \citep{Kanyongo2007}, a  $\mathcal{R} \neq 1$ affects $\mathcal{P}$ by indirectly changing  $\mathcal{W}$ as follows  

\begin{equation}
\mathcal{W}_{obs} = \mathcal{W}\sqrt{\mathcal{R}}, 
\end{equation}
where $\mathcal{R}$ is defined by the ratio of true variance ($\sigma_{\rm true}^2$)  to observed variance $(\sigma_{\rm obs}^2)$
 
\begin{equation}
\mathcal{R} =\frac{\sigma_{\rm true}^2}{\sigma_{\rm obs}^2}.
\end{equation}
By changing $\mathcal{W}$, we increase the number of observed objects necessary to reach the same  $\mathcal{P}$. For instance, considering $\mathcal{R}$ = 0.7 (0.6)   would increase the number of objects necessary to reach a power of 99.5 per cent to $\sim 150 (180)$ SN.  Therefore considering a feasible  level of purity in our data would demand $\lesssim 2\times 10^2$  high redshift SN to discriminate between a primordial Salpeter or flat IMF. 

\begin{figure*}
\centering
\includegraphics[scale=0.45]{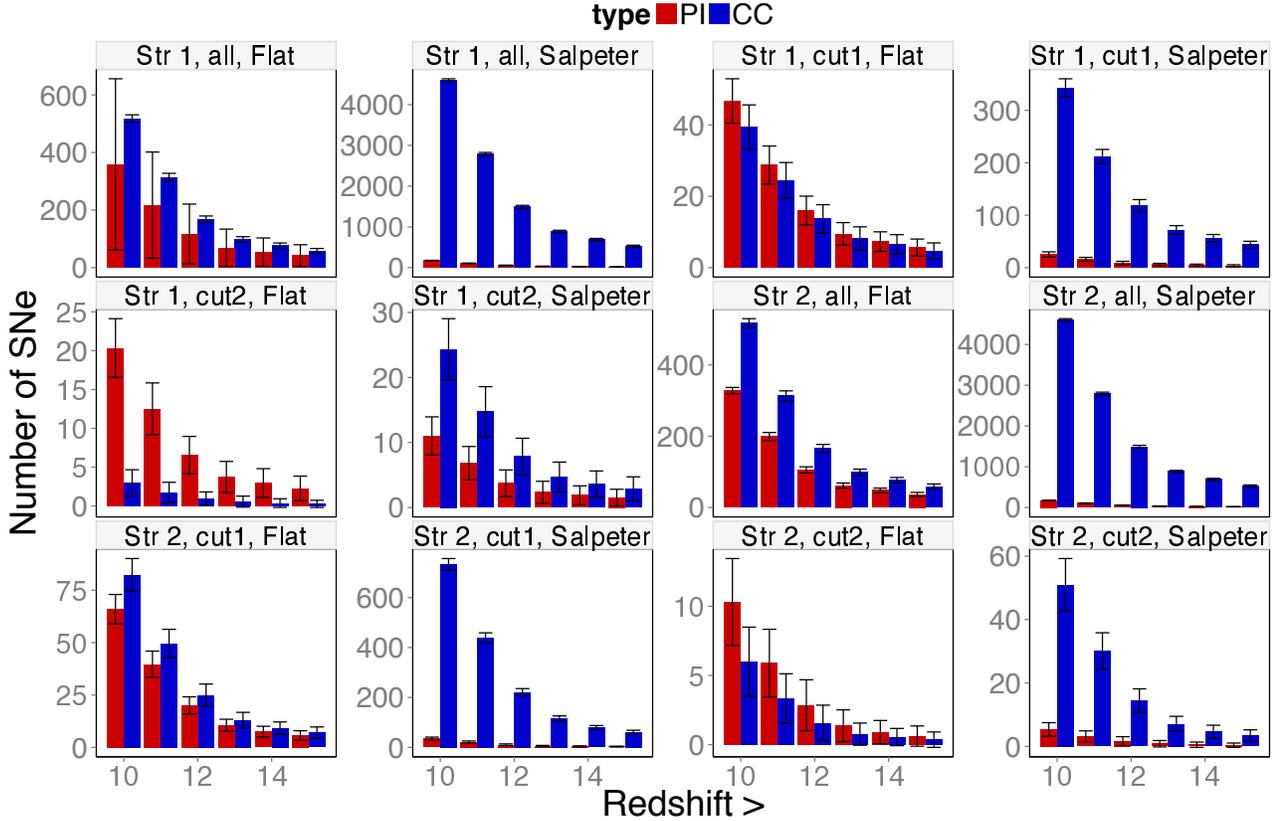}
\caption{Number of SNe detections above a certain redshift. Each column represents an average value over 200 realizations for PI (red bar ) and CC (blue bar) SN detections.  The standard deviation around the mean is  given by the error bars.  }
\label{fig:bar}
\end{figure*}

An issue    still remains,  even if we are able  to discriminate between a Pop III CC and a PI SN, we still need to distinguish between a Pop II from  Pop III SN. 
For instance,   CC SN in both populations  have similar energies and line spectra due 
to internal mixing and fallback prior to shock breakout from the surface of the star. Therefore, to be certain we are probing the Pop III IMF and not the total one,  we should observe SN in redshifts as higher as  $z \gtrsim 12$.  In Fig. \ref{fig:bar} we show the average number of detection of each type above a given redshift over 5 yrs of JWST mission. Using $\lesssim 10$ per cent of JWST time allow us to reach the minimum number to probe the IMF in few scenarios, but not all. However JWST has fuel for a 10 yrs mission, which considerably improves the possibility of achieving such goal, but only if a well planed and dedicated strategy is implemented. While the Pop III/Pop II identification has   no effect in our power analysis, it implies that, due to   a high level of contamination, it might only be possible to probe the total IMF instead of the primordial one. 

%--------------------------------------------------------------------------------------------------------------------------------------------------------------------------------------------------------------------------------%
   \begin{table}
 \label{tab:Confusion}
 \caption{Confusion Matrix.}
 \begin{center}
\begin{tabular}{lcc}
\hline
   & \multicolumn{2}{c}{Observed}\\
 \hline
 Predicted & CC & PI SN  \\
  \hline
   CC  & 302 & 129 \\

   PI SN  &   144  & 576\\
\hline  
  \end{tabular}   

 \begin{itemize}
\item Number of correct classifications: 302+576 = 878\\
\item Number of incorrect classifications: 129+144 = 273\\
\item Accuracy: 878/(878+273) $\approx$ 0.763 
\end{itemize}
\end{center}
\end{table}
%--------------------------------------------------------------------------------------------------------------------------------------------------------------------------------------------------------------------------------%

%--------------------------------------------------------------------------------------------------------------------------------------------------------------------------------------------------------------------------------%

%------------------------------------------------------------------------------------------------------------------------------------------------------------------------------------------------------------------------%

%--------------------------------------------------------------------------------------------------------------------------------------------------------------------------------------------------------------------------------%

%------------------------------------------------------------------------------------------------------------------------------------------------------------------------------------------------------------------------%

\section{Conclusions}
\label{sec:conclusions}

We have created the first, detailed synthetic observations of high-\emph{z} SNe.  With cosmological and radiation hydrodynamics simulations, together with detailed models 
of instrument properties and IGM absorption, we show that a viable survey strategy could detect up to $\sim 300$ CC and $\sim 15$ PI SNe per year.  We also find 
that information from multiple filters is crucial for 
detecting and distinguishing high-$z$ SN candidates from low-$z$ events with dropout techniques.  Our study also 
shows that the low rates of Type IIn SNe will likely prevent their detection in deep-field surveys, and all-sky missions such as \textit{Euclid} will be better suited for this task.  Survey strategies for Pop III SNe in all-sky missions such as \textit{Euclid}, {\it WFIRST} and {\it WISH} will be examined in future work.

New calculations now show that additional SN types not considered here will also be visible at high redshifts.  \citet{Whalen2013f} have now shown that 100 - 140 
$M_{\odot}$ PPI SNe will be visible to {\it JWST} at $z \sim$ 10 - 20, and, depending on the Pop III IMF, may be as numerous as PI SNe \citep[see also][]{wbh07}.  
Hypernovae, the highly energetic and asymmetric explosions of 40 - 60 $M_{\odot}$ stars, will likewise be visible out to $z \sim$ 10 - 15 \citep{Smidt2013}, and 
detection limits in redshift  for 85 - 135 $M_{\odot}$ PI SNe are now being investigated \citep{Whalen2013g}.  These additional SN classes will be considered in future
mock surveys.

Our synthetic survey allows us to perform the first comprehensive statistical study on the feasibility of using ancient  cosmic  explosions to constrain the stellar  IMF  in the early Universe by 
counting the relative detection rates of CC and PI SNe.  Given the realism with which we simulate the observing process, our synthetic sample may be easily used to 
calibrate photometric classification techniques.  Using the formalism of contingency tables together with well-established $\chi^2$ statistics, we show that at least $\sim 
10^2$ SNe detections will be necessary to distinguish between Salpeter and flat IMFs with a 99.5 per cent  confidence level.

We show that machine learning  techniques constitute a promising tool  to discern between CC and PI SNe  only relying on photometric information.  Even if the accuracy of our classification is as low as 60 per cent, we show that  $\lesssim 200$ observations  are enough to discriminate between the two IMFs. 
This represents a leap forward in high-redshift 
SN studies, which may  shed light on the long-standing puzzle of the primordial IMF. 

\section*{Acknowledgments}

D.J.W. was supported by the European Research Council under the European Community's Seventh Framework Programme (FP7/2007 - 2013) via the ERC Advanced Grant "STARLIGHT:  Formation of the First Stars" (project number 339177). 
EEOI thanks the Brazilian agencies FAPESP (2011/09525-3) and CAPES (9229-13-2) for financial support. Work at LANL was done under the auspices of the National Nuclear Security Administration of the US Department of Energy at Los Alamos National Laboratory under Contract No. DE-AC52-06NA25396. We thank Joseph Smidt for providing part of the data present in Fig. 1.  RSS thanks MPA for the hospitality during the preparation of this work. 

%------------------------------------------------------------------------------------------------------------------------------------------------------------------------------------------------------------------------%

\appendix

%------------------------------------------------------------------------------------------------------------------------------------------------------------------------------------------------------------------------%

\section{IGM Absorption}
\label{sec:IGM}

The spectra of events at $z > 6$ blueward of $1216(1+z) \rm \AA$ \citep{Gunn1965,Mesinger2004} are heavily absorbed by the neutral IGM \citep{Ciardi2011}.\footnote{In 
principle, the IGM can also absorb photons redward of $1216(1+z)\rm \AA$ by the Ly$\alpha$ damping wing cross-section \citep{Miralda1998,Mesinger2008,Bolton2013}.
This serves to lessen the discontinuity in $\tau$ at $1216(1+z)\rm \AA$, but does not affect our conclusions.}  The contribution to the optical depth $\tau_{\rm IGM}$ comes 
mainly from damped Ly$\alpha$ absorbers (DLAs), Lyman Limit Systems (LLSs), optically thin systems and resonance line scattering by the Ly$\alpha$ forest along the line 
of sight.  We account for absorption by the neutral IGM at high-$z$ by multiplying each SED by the IGM transmission function for the source redshift. The observed spectrum 
($f_{\lambda,\rm obs}$) after IGM attenuation is 
\begin{equation}
f_{\lambda,\rm obs} = f_{\lambda}e^{-\tau_{\rm IGM}}.
\end{equation}
We compute transmission through the IGM with the JAVA code \textsc{igmtransmission} \citep{Harrison2011}.  Our model uses a Monte-Carlo approach to distribute LLSs 
chosen from a redshift distribution ($dN/dz$) and an optical depth distribution ($dN/d\tau_L$), averaged over IGM 
transmission along many lines of sight \citep{Meiksin2006}. 

The contribution from optically thin systems is 
\begin{equation}\label{diffuseIGM}
\tau _{L}^{\rm IGM} = 0.07553(1+z_L)^{4.4}\left[\frac{1}{(1+z_L)^\frac{3}{2}}-\frac{1}{(1+z)^\frac{3}{2}}\right],
\end{equation}
where $z_L=\lambda /\lambda _{L}-1$ and $\lambda_{L} = 912$ \rm \AA.  The contribution due to the optically thick ($\tau_L>1$) LLSs is 
\begin{equation}\label{LLSattenuation}
\tau_{L}^{\rm LLS} = \int^{z}_{z_{L}} dz' \int^{\infty}_{1} d\tau_{L}\frac{\partial^2 N}{\partial \tau_L \partial z'}\left\{1-\exp\left[-\tau_L\left(\frac{1+z_L}{1+z'}\right)^{3} \right] 
\right \},
\end{equation}
where $\frac{\partial^2 N}{\partial \tau_L \partial z'}$ is the number of absorbers along the line of sight per unit redshift interval per unit optical depth of the system. 
\\
The spatial distribution of the LLSs is \citep{Meiksin2006}
\begin{equation}
\label{MeiksinDistribution}
\frac{dN}{dz} = N_0(1+z)^{\gamma}, 
\end{equation}
where $N_0 = 0.25$ and $\gamma = 1.5$.  The optical depth distribution is 
\begin{equation}\label{TauDistribution}
\frac{dN}{d\tau_L} \propto  \tau_L^{-\beta},  
\end{equation}
where $\beta = 1.5$.

%------------------------------------------------------------------------------------------------------------------------------------------------------------------------------------------------------------------------%

\section{Contingency tables}
\label{sec:tables}

Contingency tables are used to analyse the relationship between two or more categorical variables. It is the categorical equivalent of the scatter plot used to analyse the 
relationship between two continuous variables.  A categorical variable is one that has two or more classes, but there is no intrinsic ordering between them.  For example, 
gender is a categorical variable having two possible classes (male and female), and there is no intrinsic ordering to the categories.  Hence, the SN types can be treated 
as categorical variables since they are mutually exclusive. 

We can measure the association between the occurrence of a CC or PI SN and the IMF through their RR.    In Table B1, we have an example of RR calculation for  
the fiducial model and the case with a flat IMF, SFR10, cut2 and str2.  The RR in this case is 2.894 for CC SNe and 0.053 for PI SNe, meaning that a Salpeter IMF is 
$\sim 3$ times more likely to produce a CC event than in the flat IMF.

   \begin{table}
 \label{tab:cut2_str2_sf}
 \caption{Contingency table for the fiducial model  and  the case with a  flat  IMF,  SFR10, cut2 and  str2. }
 \begin{center}
\begin{tabular}{lcc|cc}
\hline
 &  \multicolumn{2}{c}{SN type}\\
 \hline
  IMF & PI SN & CC & Total by IMF\\
  \hline
  $\rm Salpeter_{(all)}$ & 339 (359.77) & 9312 (9291.23)  & 9651\\

  $\rm Flat_{(cut2;str2)}$ &   22 (1.23) & 11 (31.77) &33\\
\hline  
  Total by type & 361 & 9323 & 9684\\
  \hline
 \end{tabular}

  \begin{itemize}
\item Relative risk: PI SN = $\frac{339/9651}{22/33}  \approx 0.053$\quad CC = $\frac{9312/9651}{11/33}  \approx 2.894$
\item $\chi^2 = 365.5$ \quad  p-value $< 0.0005$

\end{itemize}

  \end{center}
 \end{table}

\begin{table}
 \label{tab:cut1_str1_sf}
 \caption{Contingency table for the fiducial model and  the case with a  flat  IMF, SFR10, cut2 and str1. }
 \begin{center}
\begin{tabular}{lcc|cc}
\hline
 &  \multicolumn{2}{c}{SN type}\\
 \hline
  IMF & PI SN & CC & Total by IMF\\
  \hline
  $\rm Salpeter_{(all)}$ & 339 (409.19) & 9312 (9241.81)  & 9651\\

  $\rm Flat_{(cut1;str1)}$ &   76 (5.80) & 61 (131.19) &137\\
\hline  
  Total by type & 415 & 9373 & 9788\\
  \hline
    \end{tabular}

  \begin{itemize}

\item Relative risk: PI SN = $\frac{339/9651}{76/137}  \approx 0.063$\quad CC = $\frac{9312/9651}{61/137}  \approx 2.167$
\item $\chi^2 = 898.3$ \quad  p-value $< 0.0005$

\end{itemize}
  \end{center}
 \end{table}

%--------------------------------------------------------------------------------------------------------------------------------------------------------------------------------------------------------------------------------%
\begin{table}
 \caption{Contingency table for the fiducial model  and  the case with a  flat  IMF,  SFR10, cut2, and str1. }
 \begin{center}
 \label{tab:cut2_str1_sf}
\begin{tabular}{lcc|cc}
\hline
 &  \multicolumn{2}{c}{SN type}\\
 \hline
  IMF & PI SN & CC & Total by IMF\\
  \hline
  $\rm Salpeter_{(all)}$ & 339 (371.54) & 9312 (9279.46)  & 9651\\

  $\rm Flat_{(cut2;str1)}$ &   34 (1.46) & 4 (36.54) &38\\
\hline  
  Total by type & 373 & 9316 & 9689\\
  \hline
    \end{tabular}

  \begin{itemize}

\item Relative risk: PI SN  $\frac{339/9651}{34/38}  \approx 0.039$\quad CC = $\frac{9312/9651}{4/38}  \approx 9.166$
\item $\chi^2 = 755.6$ \quad  p-value $< 0.0005$

\end{itemize}
  \end{center}
 \end{table}
%--------------------------------------------------------------------------------------------------------------------------------------------------------------------------------------------------------------------------------%

  \begin{table}

 \caption{Contingency table for the fiducial model and  the case with a   Salpeter IMF, SFR10, cut2, and str2.  }
 \begin{center}
 \label{tab:cut2_str2_ss}
\begin{tabular}{lcc|cc}
\hline
 &  \multicolumn{2}{c}{SN type}\\
 \hline
  IMF & PI SN & CC & Total by IMF\\
  \hline
  $\rm Salpeter_{(all)}$ & 339 (344.04) & 9312 (9306.96)  & 9651\\

  $\rm Salpeter_{(cut2;str2)}$ &   9 (3.96) &102 (107.04) &111\\
\hline  
  Total by type & 348 & 9414 & 9762\\
 
  \hline
    \end{tabular}

  \begin{itemize}
  \item Relative risk: PI SN = $\frac{339/9651}{9/111}  \approx  0.433$\quad CC = $\frac{9312/9651}{102/111}  \approx 1.050$
   \item $\chi^2 = 6.74$ \quad  p-value $= 0.02$

\end{itemize}
  \end{center}
   \label{tab:ap1}
 \end{table}

 \begin{table}

 \caption{Contingency table for the fiducial model  and the case with   a  Salpeter IMF, SFR10, cut1, and str1.  }
 \begin{center}
 \label{tab:cut1_str1_ss}
\begin{tabular}{lcc|cc}
\hline
 &  \multicolumn{2}{c}{SN type}\\
 \hline
  IMF & PI SN & CC & Total by IMF\\
  \hline
  $\rm Salpeter_{(all)}$ & 339 (356.78) & 9312 (9294.22)  & 9651\\

  $\rm Salpeter_{(cut1;str1)}$ &   39 (21.22) &535 (552.78) &574\\
\hline  
  Total by type & 378 & 9847 & 10225\\
 
  \hline
    \end{tabular}

  \begin{itemize}
  \item Relative risk: PI SN = $\frac{339/9651}{39/574}  \approx  0.517$\quad CC = $\frac{9312/9651}{535/574}  \approx 1.035$
   \item $\chi^2 = 16.39$ \quad  p-value $= 0.0005$

\end{itemize}
  \end{center}
 \end{table}

  \begin{table}
 \label{tab:cut2_str1_ss}
 \caption{Contingency table for the fiducial model  and  the case with  a  Salpeter IMF, SFR10, cut2, and str1.  }
 \begin{center}
\begin{tabular}{lcc|cc}
\hline
 &  \multicolumn{2}{c}{SN type}\\
 \hline
  IMF & PI SN & CC & Total by IMF\\
  \hline
  $\rm Salpeter_{(all)}$ & 339 (354.01) & 9312 (9296.98)  & 9651\\

  $\rm Salpeter_{(cut2;str1)}$ &   17 (1.98) &37 (52.01) &54\\
\hline  
  Total by type & 356 & 9349 & 9705\\
 
  \hline
    \end{tabular}

  \begin{itemize}
  \item Relative risk: PI SN = $\frac{339/9651}{17/54}  \approx  0.112$\quad CC = $\frac{9312/9651}{37/54}  \approx 1.408$
   \item $\chi^2 = 118.9$ \quad  p-value $= 0.0005$

\end{itemize}
  \end{center}
 \end{table}
%------------------------------------------------------------------------------------------------------------------------------------------------------------------------------------------------------------------------%

\section{Noise calculation}
\label{sec:noise}

This section briefly describes how the uncertainties in measured fluxes and magnitudes are built (error bars in Fig. \ref{fig:LCs}).  A more in-depth discussion and example 
calculation can be found in the SNANA Manual,\footnote{http://sdssdp62.fnal.gov/sdsssn/SNANA-PUBLIC/doc/ snana\_manual.pdf} section 4.12.

Consider the simulation of the observation of an event in a filter at one epoch. The SED describes the evolution of the spectrum of this event over time in the frame of the 
SN.  This information is then redshifted into the observer frame, taking into account cosmological dimming, absorption by the IGM, Milky Way extinction, the collection area
of the telescope mirror, total exposure time, etc.  The product of this calculation is the AB magnitude ($m$) of the SN, minus the errors due to the observing process.  In 
order to evaluate these errors and account for them into our synthetic observations, we need to incorporate at least 3 important sources of uncertainty:

\begin{itemize}
\item the determination of the signal itself, $\sigma_{\rm sig}$;
\item sky background, $\sigma_{\rm sky}$;
\item calibration process (determination of the zero point), $\sigma_{\tt ZPT}$.
\end{itemize}

The measured source signal can be expressed in terms of the number of photoelectrons generated in the CCD, $N_{\rm pe}$.  In this case, 
 $\sigma_{\rm sig}=G^{-1}\sqrt{N_{\rm pe}}$ in analog-to-digital-units (ADU), where $G$ is the CCD gain (pe/ADU).  The sky background error is given by 
$\sigma_{\rm sky}=\pm \sqrt{A \sigma_{\rm skypix}^2}$, where $A$ is the PSF equivalent area, represented by a simple Gaussian of full-width-at-half-maximum 
$\sigma_1 (A=4\pi\sigma_1^2)$, and $\sigma_{\rm skypix}$ is the measured uncertainty in sky brightness per pixel (ADU$/$pixel).  The error coming from the zero point 
magnitude ($\sigma_{\tt ZPT}$) encodes information about the atmospheric transparency and telescope aperture and efficiency, but does not depend on the signal from 
this particular object.  Thus, the statistical error due to photon statistics is composed of $\sigma_{\tt sig}/F$ and $\sigma_{\tt sky}/F$.  These should be added in 
quadrature to $\sigma_{\tt ZPT}$, which represents an additional statistical error coming from stellar calibration. 

Finally, the uncertainty in flux determination (in ADU) is given by
\begin{eqnarray}
 \sigma_{ F}&=& F\times\sqrt{\frac{\left(\sigma_{\rm sig}^2 + \sigma_{\rm sky}^2\right)}{F^2} + \sigma_{\tt ZPT}^2 },
\end{eqnarray} 
where ${F}=10^{-0.4(m - m_{\tt ZPT})}$ is the flux from the source, $m$ is its corresponding magnitude, and $m_{\tt ZPT}$ is the zero point magnitude.  To calculate the 
error in $m$, we define
\begin{equation}
\sigma_{ m\pm} = -5\log_{10}\left(\frac{ F}{{ F} \pm \sigma_{ F}}\right).
\end{equation}
This equation is used to determine right and left limits for the case of asymmetric errors.  In the simple case where the confidence levels in magnitude are symmetric, the 
observed error  is defined as 
\begin{equation}
\sigma_{ m}= \frac{|\sigma_{ m+}| + |\sigma_{ m-}|}{2}.
\end{equation} 
%

%------------------------------------------------------------------------------------------------------------------------------------------------------------------------------------------------------------------------%

\section{JWST Specifications}
\label{sec:JWST}

To account for instrument characteristics, we use specifications from NIRCam and other sources shown in Table \ref{tab:t_I}. For completeness, we also show explicit 
values for the zero point, sky magnitude and error in sky magnitudes for each filter in Table \ref{tab:calc_input}. The CCD readout noise and sky-noise are determined 
by the CDD-noise and $\sigma_{\rm sky}$ parameters summed in quadrature over A based on the PSF fitting. 

%------------------------------------------------------------------------------------------------------------------------------------------------------------------------------------------------------------------------%

\begin{table}
\caption{\textit{JWST} technical specifications used to construct the simulation library.  ZPTSIG: additional smearing to zero; full-width-at-half-maximum (FWHM), 
NIRCam field of view (FOV), mirror collecting area (A), pixel scale and CCD gain/noise \citep[][and references therein]{Gardner2006}.}
\centering
\begin{tabular}{ccc}
\hline
Feature & value\\
\hline
CCD gain (e$^-$/ADU)    &  3.5\\
CCD  noise (e$^-$/pixel)   & 4.41\\
Pixel scale (arcsec/pixel)  &   0.032  for F070W-F200W \\
 & 0.065 for F277W-F444W\\
FWHM (pixels) & 2  \\
$\sigma_{\tt ZPT}$ & 0.004 \\
FOV ($\rm arcmin^2$)   & 9.68        \\       
A (cm$^2$) & 2.5$\times 10^5$ \\
\hline
\label{tab:t_I}
\end{tabular}
\end{table}
%------------------------------------------------------------------------------------------------------------------------------------------------------------------------------------------------------------------------%

%------------------------------------------------------------------------------------------------------------------------------------------------------------------------------------------------------------------------%

\begin{table}
\caption{Inputs used in the construction of the SNANA simulation library (SIMLIB) file for \textit{JWST}.  Columns correspond to NIRCam filter, zero point for AB 
magnitudes ($m_{\rm ZPT}$), sky brightness ($m_{\rm sky}$) and error in sky brightness ($\rm \sigma_{sky}$). All values were calculated for an exposure time 
of $10^3$s.}
\centering
\begin{tabular}{ l c c  c }
\hline
NIRCam  & $m_{\tt ZPT}$ & $m_{\tt sky}$  & $\sigma_{\tt sky}$\\
filter 	& (mag/arcsec$^{\rm \scriptsize{2}}$) & (mag/arcsec$^{\rm \scriptsize{2}}$) & (ADU/pixel) \\
 \hline 
F070W & 27.87 & 27.08 & 1.24    \\
 F090W &28.41 &   26.90 & 1.34  \\
F115W &  28.89 &   26.76 &  1.39   \\
F150W &  29.51 &  26.88  & 1.35  \\
F200W & 30.20 & 26.96  & 1.34   \\
F277W &  30.85 &  26.32  & 1.75  \\
F356W &  31.38 &  26.75  &  1.43    \\
F444W &  31.88 &  25.58  & 2.47   \\
\hline
\end{tabular}
\label{tab:calc_input}
\end{table}
%------------------------------------------------------------------------------------------------------------------------------------------------------------------------------------------------------------------------%

%------------------------------------------------------------------------------------------------------------------------------------------------------------------------------------------------------------------------%
\section{Supernova photometric classification}

\subsection{Kernel Principal Components Analysis}
\label{sec:kpca}
KPCA generalizes the standard PCA by first mapping the data into a higher dimensional feature space $\mathbb{F}$:

\begin{eqnarray}
\Phi : \mathbb{R}^n &\rightarrow& \mathbb{F}\nonumber\\
\mathbf{x}&\rightarrow& \Phi(\mathbf{x}),
\end{eqnarray}
where $\Phi$ is a nonlinear function and $\mathbb{F}$ has arbitrary number of dimensions.

The covariance matrix,  $C_{\rm F} \in \mathbb{F}$,  will be defined similarly as 
\begin{equation}
C_{\rm F} = \frac{1}{N}\sum_{i=1}^N\Phi(x_i)\Phi(x_i)^T.
\label{eq:kcovmatrix}
\end{equation}
We assume that $\Phi(\mathbf{x}_i)$ are centered in feature space. 
Consider $\mathbf{v}_{\Phi}^l$ the $l-th$ eigenvector of $C_{\rm F}$ and $\lambda_{\Phi}^l$ its $l-th$ eigenvalue.
We can define a kernel $N\times N$ matrix 
\begin{equation}
K_F(\mathbf{x}_i,\mathbf{x}_j) = (\Phi(\mathbf{x}_i)\cdot\Phi(\mathbf{x}_j)), 
\end{equation}
which allows us to compute the value of dot product in $\mathbb{F}$ without carrying out the map $\Phi$.  The kernel function has to satisfy the Mercer's theorem to ensure that it is possible to construct a mapping into a space where $K_{\rm F}$ acts as a dot product. The projection of a new test point, $\mathbf{n}$, is given by
\begin{equation}
(\mathbf{v}_{\Phi}^l\cdot\Phi(\mathbf{n}))=\sum_{i=1}^N\alpha_{\Phi_i}^l K_F(\mathbf{x_i},\mathbf{n}),
\label{eq:featproj2}
\end{equation}
where $\alpha_{\Phi_i}^l$ is defined by the solutions to the eigenvalue equation $N\lambda_{\Phi}\alpha_{\Phi} = K_F\alpha_{\Phi}$.

It is important to stress that all the arguments shown in this appendix rely on the assumption that the data are centered in feature space.   Where the  centered kernel matrix, $\widetilde{K_F}$, can be expressed in terms of the non-centered kernel matrix, $K_F$, as
\begin{equation}
\widetilde{K_F}=K_F-1_N K_F-K_F 1_N+1_M K_F 1_N,
\label{eq:KFcent}
\end{equation}
where $(1_N)_{ij}=1/N$. The reader should be aware that we always refer to the centered kernel matrix $\widetilde{K_F}$. However, for the sake of simplicity, the tilde is not used in our notation.

Finally,  we  need to choose a form for the kernel function $k(\mathbf{x}_i,\mathbf{x}_j):=K_{{F}_{ij}}$. 
For the sake of simplicity, we make an \textit{a priori} choice of using  a Gaussian kernel, 
\begin{equation}
k(\mathbf{x}_i,\mathbf{x}_j) = \exp\left[-\frac{\|\mathbf{x}_i-\mathbf{x}_j\|^2}{2\sigma^2}\right],
\label{eq:kernel}
\end{equation}
where the value of $\sigma$ is determined by a cross-validation process.

%------------------------------------------------------------------------------------------------------------------------------------------------------------------------------------------------------------------------%
%
\subsection{The $k$-Nearest Neighbor algorithm} 
\label{subsec:kNN}
The kNN  begins with the training sample organized as $q_i=(x_i,y_i)$, where $x_i$ is the $i-th$ data vector and $y_i$ its label, and a definition of distance between 2 data vectors $d(x_i,x_j)$. Given a new unlabelled test point $q_t(x_t, )$, the algorithm computes the distance between $x_t$ and all the other points in the training sample, $d(x_t,\mathbf{x})$, ordering them from lower to higher distance. The labels of the first $k$ data vectors (the ones closer to $x_t$) are counted as votes in the definition of $y_t$. Finally, $y_t$ is set as the label with highest number of votes. Throughout our analysis, we used an Euclidean distance metric and order $k=1$. 

%%------------------------------------------------------------------------------------------------------------------------------------------------------------------------------------------------------------------------%
%
\subsection{Cross-validation}
\label{ap:CV}
The main idea behind the cross-validation procedure is to remove a random set of $M$ data points, $T^{\rm{out}}$ from the sample. The remaining part  is given as input  for a given classifier algorithm   and used to classify the points in $T^{\rm{out}}$. In this way, we can measure the success rate of the classifier over different random choices of $T^{\rm{out}}$ and also compare results from different classifiers given the same training and $T^{\rm{out}}$ sets \citep[see e.g., ][]{Arlot2010}.
\\
The the number of points in $T^{\rm{out}}$ is a free parameter and  must be defined based on the clustering characteristics of the given data set. Here we chose the most classical exhaustive data splitting procedure, sometimes called \textit{Leave One Out} (LOO) algorithm. As the name states, we construct $N$ sub-samples $T^{\rm{out}}$, each one containing only one data point, $M=1$. The training sample is then cross-validated and the performance judged by the average number of correct classifications. 
\\
Data exhaustive algorithms like LOO have a larger variance in the final results, although, they are highly recommended for avoiding biases regarding local data clustering and some non-uniform geometrical distribution of data points in a given parameter space. 

%------------------------------------------------------------------------------------------------------------------------------------------------------------------------------------------------------------------------%

\footnotesize{

}

\label{lastpage}
\end{document}